\documentclass[aps,prb,amsmath,superscriptaddress,amssymb,footinbib,10pt,twocolumn,longbibliography]{revtex4-2} 
\usepackage[colorlinks=true,linkcolor=blue,citecolor=blue,urlcolor=blue]{hyperref}
\usepackage{graphicx}
\usepackage{amsmath}
\usepackage{bm} 
\usepackage{xcolor}
\usepackage[version=4]{mhchem}
\usepackage{braket}

\renewcommand{\d}{\text{d}}
\newcommand{\D}{\mathcal{D}}
\DeclareMathOperator{\tr}{tr}

\begin{document}

\title{Superconductivity in kagome metals due to soft loop-current fluctuations}

\author{Daniel J. Schultz}
\thanks{Corresponding author: Daniel J. Schultz, email: \href{mailto:daniel.schultz@kit.edu}{daniel.schultz@kit.edu}}
\affiliation{Institute for Theoretical Condensed Matter Physics, Karlsruhe Institute of Technology, 76131 Karlsruhe, Germany}

\author{Grgur Palle}
\affiliation{Department of Physics, The Grainger College of Engineering, University of Illinois Urbana-Champaign, Urbana, Illinois 61801, USA}
\affiliation{Anthony J. Leggett Institute for Condensed Matter Theory, The Grainger College of Engineering, University of Illinois Urbana-Champaign, Urbana, Illinois 61801, USA}

\author{Asimpunya Mitra}
\affiliation{Department of Physics, University of Toronto, Toronto, Ontario M5S 1A7, Canada}

\author{Yong Baek Kim}
\affiliation{Department of Physics, University of Toronto, Toronto, Ontario M5S 1A7, Canada}

\author{Rafael M. Fernandes}
\affiliation{Department of Physics, The Grainger College of Engineering, University of Illinois Urbana-Champaign, Urbana, Illinois 61801, USA}
\affiliation{Anthony J. Leggett Institute for Condensed Matter Theory, The Grainger College of Engineering, University of Illinois Urbana-Champaign, Urbana, Illinois 61801, USA}

\author{J\"org Schmalian}
\affiliation{Institute for Theoretical Condensed Matter Physics, Karlsruhe Institute of Technology, 76131 Karlsruhe, Germany}
\affiliation{Institute for Quantum Materials and Technologies, Karlsruhe Institute of Technology, 76131 Karlsruhe, Germany}

\date{\today}

\begin{abstract}
\textbf{Abstract:} We demonstrate that soft fluctuations of translation symmetry-breaking loop currents provide a mechanism for unconventional superconductivity in kagome metals that naturally addresses the multiple superconducting phases  observed under pressure. Focusing on the rich multi-orbital character of these systems, we show that loop currents involving both vanadium and antimony orbitals generate low-energy collective modes that couple efficiently to electrons near the Fermi surface and mediate attractive interactions in two distinct unconventional pairing channels. While loop-current fluctuations confined to vanadium orbitals favor chiral $d+id$ superconductivity, which spontaneously breaks time-reversal symmetry, the inclusion of antimony orbitals stabilizes an $s^{\pm}$ state that is robust against disorder. We argue that these two states are realized experimentally as pressure increases and the antimony-dominated Fermi surface sheet undergoes a Lifshitz transition.
\end{abstract}

\maketitle

\noindent{\large\textbf{Introduction}}

\noindent Kagome metals of the family \ce{AV3Sb5}, with A = Cs, Rb, and K, exhibit a rich phase diagram featuring charge density-wave (CDW) order and superconductivity (SC) as well-established symmetry-broken states \cite{ortiz_cs_2020,ortiz_new_2019,wilson_av3sb5_2024,luo_unique_2023,xu_v3_2025, zhao_cascade_2021, wu_charge_2022,ge_charge-_2024, liu_charge-density-wave-induced_2021, lou_charge-density-wave-induced_2022, xiao_coexistence_2023, ratcliff_coherent_2021, frachet_colossal_2024, li_discovery_2022, ye_distinct_2024, nguyen_electronic_2022, asaba_evidence_2024,miao_geometry_2021, jiang_kagome_2023, duan_nodeless_2021, wang_quantum_2023, mu_s-wave_2021, liege_search_2024,ortiz_superconductivity_2021}.
Moreover, they have emerged as prime candidates for hosting loop-current (LC) states that break both time-reversal symmetry and translational invariance,
suggesting that one phase may facilitate the emergence of the other (for a recent review, see \cite{fernandes_loop-current_2025}).
In this work, we demonstrate that fluctuations of loop currents or, equivalently, the fluctuations of orbital-magnetic fluxes, give rise to unconventional superconductivity in the kagome metals.

The nature of the CDW-induced lattice distortions has largely been clarified through scanning tunneling microscopy \cite{shumiya_intrinsic_2021, zhao_cascade_2021,jiang_unconventional_2021,li_no_2022} and X-ray scattering experiments \cite{chen_charge_2022, ortiz_cs_2020, li_observation_2021, ortiz_fermi_2021, elmers_chirality_2025, subires_order-disorder_2023, stier_pressure-dependent_2024, scagnoli_resonant_2024, li_rotation_2022, stahl_temperature-driven_2022}. Theoretically, they have been proposed to arise from the interplay between phonons and electronic states which are primarily made of vanadium $3d$ orbitals, with additional contributions coming from the antimony $5p$ states \cite{tan_charge_2021,christensen_theory_2021,ritz_impact_2023}.  The LC state, on the other hand, has often been tied to the Van Hove singularities near the Fermi energy whose states are dominated by vanadium $3d$ orbitals \cite{kiesel_sublattice_2012,kiesel_unconventional_2013,denner_analysis_2021,park_electronic_2021,lin_complex_2021,feng_low-energy_2021,christensen_loop_2022,tazai_charge-loop_2023,wagner_phenomenology_2023,yang_intertwining_2023,dong_loop-current_2023, li_intertwined_2024,fu_exotic_2024}. Interestingly, SC seems to be closely related to a Fermi surface sheet dominated by antimony $5p$ states, as its suppression via a Lifshitz transition coincides with a sudden drop in the transition temperature \cite{tsirlin_effect_2023,oey_fermi_2022}. Given the distinct role of these orbital states, the link between CDW and LC states, on the one hand, and pairing, on the other, is an open problem.
One possible answer is that CDWs and SCs are both driven by strong electron-phonon interactions \cite{ritz_superconductivity_2023, xie_electron-phonon_2022,luo_electronic_2022, uykur_optical_2022}.
Observations that support some form of time-reversal symmetry-breaking in the SC state \cite{guguchia_tunable_2023,deng_evidence_2024} and the fact that electron-phonon interactions do not tend to support sufficiently strong LC states, however, suggests that purely electronic interactions are important.

There are robust experimental reports of time-reversal symmetry-breaking inside the CDW state \cite{jiang_unconventional_2021,mielke_time-reversal_2022,guo_switchable_2022,xu_three-state_2022,guguchia_tunable_2023,xing_optical_2024,asaba_evidence_2024,guo_correlated_2024,deng_chiral_2024,gui_probing_2025}, although a net magnetization is unlikely to be present \cite{saykin_high_2023}. Nevertheless, it is unclear whether LCs form a stable long-range order at zero magnetic field, or whether they remain fluctuating \cite{graham_depth-dependent_2024} and condense only in the presence of a magnetic field, with an onset that either coincides with the CDW transition or occurs at a lower temperature \cite{guo_correlated_2024}.
Either way, this broad set of experimental observations strongly supports a scenario in which LCs are present as soft collective excitations.
Their fluctuations should then play a role in the low-energy physics of the kagome metals.
The goal of this paper is to analyze the consequences of these fluctuations.
As we shall see, this requires detailed knowledge of the complex electronic structure.

\begin{figure*}[t]
\centering
\includegraphics[width=\textwidth]{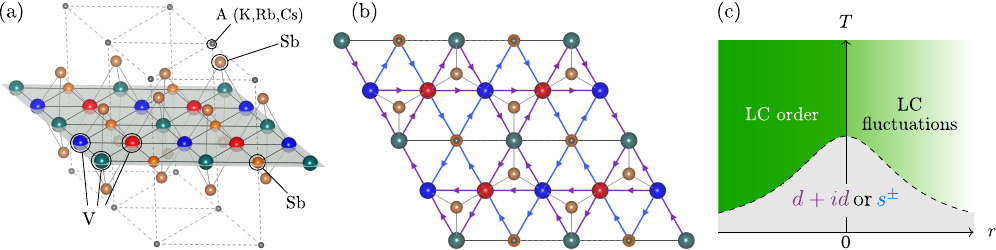}
\caption{\textit{Crystal structure and summary of main result.} (a) The crystal structure of the \ce{AV3Sb5} kagome metals. (b) Top-down view of the crystal structure, showing the V-kagome plane (red, blue, teal), the triangular lattice of planar Sb atoms (orange; below small gray circles), the  honeycomb lattice of (brown) apical Sb (which live both above and below the V-kagome plane), and the triangular lattice of the alkali metal A (K,Rb,Cs) which live directly above the in-plane Sb. The atom sizes are not to scale and (K,Rb,Cs) have been shrunk to appear smaller than Sb for visualization purposes. (c) A schematic phase diagram of our envisioned scenario. $T$ is temperature and $r$ is a tuning parameter (e.g., pressure). For $r > 0$ the system is disordered, yet loop-current fluctuations can still be soft and mediate Cooper pairing. It is this superconductivity that we are studying in the current work. V-V loop currents (purple arrows in panel (b)) drive $d+id$ pairing, whereas V-Sb loop currents (blue arrows in panel (b)) drive $s^{\pm}$ pairing.} \label{fig:crystal_structure}
\end{figure*}

\begin{figure}[t]
\centering
\includegraphics[width=\columnwidth]{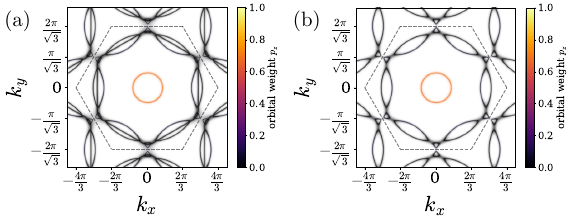}
\caption{\textit{Fermi surface of realistic tight-binding models.} The Fermi surfaces of (a) the 30 band model and (b) the 13 band model, consisting only of orbitals (or linear combinations thereof) which are odd under $\sigma_h\colon z \mapsto -z$. The key difference between (a) and (b) is that (a) has an additional large hexagonal Fermi surface whose vertices almost reach the $\mathbf{M}$-points in the Brillouin zone which is due to orbitals even under $\sigma_h$. The remaining features of importance are the Van Hove singularities at the $\mathbf{M}$-points, which the Fermi surface comes close to. The electrons near the $\mathbf{M}$-points are primarily made of $d_{zx},d_{yz}$ vanadium orbitals. There is also a circular pocket around the $\mathbf{\Gamma}$-point, arising primarily from $p_z$ orbitals of planar Sb atoms (orange shading).} \label{fig:fermi_surface}
\end{figure}

\begin{figure*}[t]
\centering
\includegraphics[width=\textwidth]{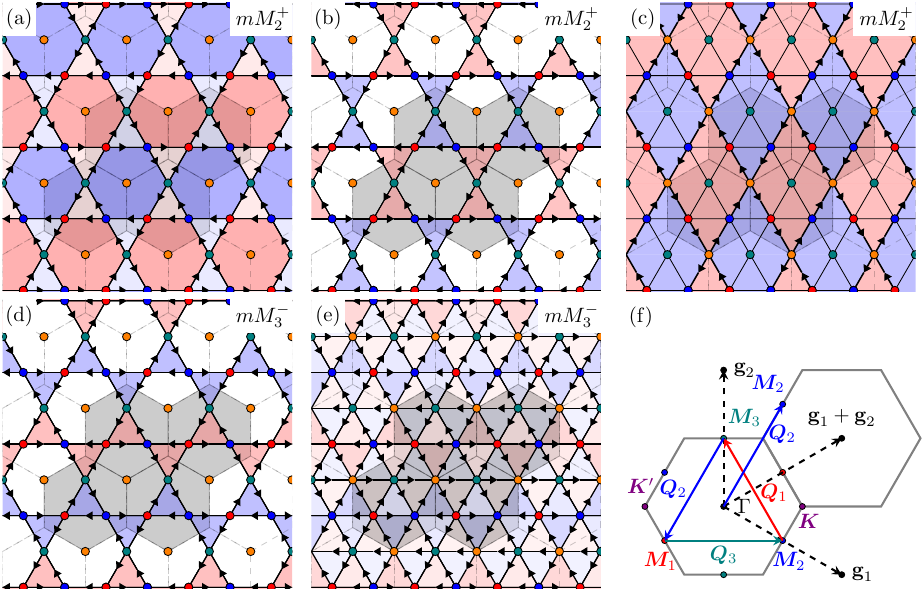}
\caption{\textit{Loop current patterns and Brillouin zone.} The 5 LC pattern possibilities that we study (a)--(e) and a sketch of the Brillouin zone (f). Each pattern comes in a set of 3, with ordering wave vectors $\mathbf{M}_{\ell=1,2,3}$. Only the representative with ordering wave vector $\mathbf{M}_3$ is shown. The other two may be obtained via $C_{6z}$ rotations.
The red, blue, and teal sites are V and they form a kagome lattice. The orange sites are the planar Sb and they form a triangular lattice and live in the centers of the hexagons formed by the kagome lattice. This reflects the same color scheme as in Fig.~\ref{fig:crystal_structure}. The greyed-out region is the extended $2\times 2$ unit cell that our LC patterns live in. The red and blue shadings denote whether the flux of a given plaquette is out-of-page or in-page, respectively.
The (a)--(c) LC patters belong to the (even-parity) $mM_{2}^+$ irrep, while (d)--(e) belong to the (odd-parity) $mM_{3}^-$ irrep; see Supplementary, Sec.~1.
(a),(b),(d) have only V-V currents, while (c),(e) have in addition V-Sb currents.
Note that the ordering wave vectors $\mathbf{Q}_{\ell}$, which connects different $\mathbf{M}$-points, are themselves $\mathbf{M}$-points (e.g., $\mathbf{Q}_3 = \mathbf{M}_2 - \mathbf{M}_1 = \mathbf{M}_3 = - \mathbf{M}_3$ up to a reciprocal lattice vector). The color scheme for these three $\mathbf{M}$-vectors (red, blue, teal) is chosen to coincide with the sublattice weight of the kagome sites (also red, blue, teal) at different $\mathbf{M}$-points in the Brillouin zone.} \label{fig:LCO_patterns}
\end{figure*}

The electronic structure of the \ce{AV3Sb5} systems consists of multiple bands made up of vanadium (V) and antimony (Sb) states \cite{cai_angle-resolved_2024, hu_electronic_2023, bhandari_first-principles_2024, ortiz_cs_2020, kang_twofold_2022,nakayama_multiple_2021}; see Figs.~\ref{fig:crystal_structure},\ref{fig:fermi_surface}. The V $3d$ states play an important role near the three saddle points $\mathbf{M}_\ell$, $\ell \in \{1,2,3\}$. These points are connected by three wave vectors $\mathbf{Q}_\ell$ that coincide with the in-plane components of the CDW modulation (e.g., $\mathbf{Q}_{3} = \mathbf{M}_2 - \mathbf{M}_1$; see Fig.~\ref{fig:LCO_patterns}(f)), highlighting the crucial role played by these electronic degrees of freedom in the formation of the CDW state, in combination with lattice degrees of freedom. Sb states, frequently ignored in low-energy models of the kagome metals, are important in making the $\mathbf{M}$-point saddle points cross the Fermi level at non-zero $k_z$  \cite{ritz_superconductivity_2023} and also form a separate Fermi surface sheet near the $\mathbf{\Gamma}$-point. The states from the apical and planar Sb atoms were recently shown to be significant in microscopic models of LC and CDW order \cite{jeong_crucial_2022, li_intertwined_2024,ritz_impact_2023}. Moreover, they are also closely tied to superconductivity since the pressure-induced suppression of this antimony Fermi surface sheet via a Lifshitz transition \cite{tsirlin_effect_2023,bhandari_pressure_2025, si_charge_2022} coincides with the disappearance of superconductivity \cite{zhang_pressure-induced_2021, du_pressure-induced_2021}. Interestingly, another superconducting state is then observed at higher pressure \cite{zhang_pressure-induced_2021}, even without this Sb-Fermi pocket. In addition, quasi-particle interference spectroscopy of \ce{KV3Sb5} finds evidence that the superconducting gap on the Sb-Fermi surface sheet is comparatively large \cite{deng_chiral_2024}. This suggests that, at least at low pressure, Sb states are important in determining the superconducting properties. Given the complexity of the electronic structure and the nature of competing states, a theory for the mechanism of superconductivity must go beyond simple model descriptions and properly account for the symmetry and the microscopic nature of symmetry-broken and fluctuating states.

In this paper, we demonstrate that LC fluctuations give rise to two dominant unconventional superconducting states depending on the system's paramaters: $s^{\pm}$ pairing with a large SC gap at the Sb-Fermi surface, and a chiral $d+id$ state that breaks time-reversal symmetry. These results follow from a microscopic model that fully incorporates the multi-orbital character of the kagome lattice, including the 30 V-$3d$ and Sb-$5p$ orbitals per unit cell \cite{li_origin_2023}.
We then carry out a comprehensive symmetry classification of LC states, identifying allowed LC patterns consistent with the lattice and time-reversal symmetry-breaking.
Given the crucial role of Sb states in shaping low-energy fluctuations, we extend the conventional treatment, which considers LCs confined to V sites, to include LC patterns that traverse between V and Sb orbitals. The key result is that the nature of the SC state is determined by which of these two types of symmetry-equivalent LC patterns display the strongest fluctuations.
LCs involving only V orbitals favor $d+id$ pairing, whereas LC involving both V and Sb orbitals favor $s^{\pm}$ pairing. We identify these two distinct pairing states with the two separate SC states observed experimentally in \ce{CsV3Sb5} as pressure is increased \cite{zhang_pressure-induced_2021}.
To avoid complications related to the reconstruction of the Fermi surface in the CDW/LC ordered state, we focus on the regime without long-range order.
This corresponds to the $r > 0$ region of Fig.~\ref{fig:crystal_structure}(c), which can be experimentally achieved
by suppressing the long-range CDW/LC order with pressure \cite{tsirlin_effect_2023, chen_double_2021, zhu_double-dome_2022,chen_highly_2021, stier_pressure-dependent_2024, qian_revealing_2021, yu_unusual_2021} or doping \cite{oey_fermi_2022, kautzsch_incommensurate_2023, labollita_tuning_2021}.
LC fluctuations have also been proposed as a pairing glue in the cuprates \cite{varma_theory_2006, varma_non-fermi-liquid_1997, simon_detection_2002, allais_loop_2012, lederer_observable_2012}. Recent work \cite{shi_loop_2023,palle_superconductivity_2024}, which shows that they are not effective at driving superconductivity, appears at first glance to contradict our results. However, as we explain later, the crucial distinction that allows soft LC fluctuations to efficiently induce Cooper pairing in the kagome systems is that they break translation symmetry, in which case the reservations of Refs.~\cite{shi_loop_2023,palle_superconductivity_2024} no longer apply.

\vspace{6pt}
\noindent{\large\textbf{Results}}

\noindent\textbf{Electronic structure:} To describe the electronic structure of \ce{AV3Sb5}, we use the $30$-band tight-binding model of Ref.~\cite{li_origin_2023}. This model includes five V-$3d$ states at the three kagome sublattices in the unit cell and three Sb-$5p$ states at five locations. In our analysis, we focus on the two-dimensional states with $k_z=0$, thereby accounting for the anisotropic electronic structure of \ce{AV3Sb5}. We may then split the orbitals into those that are even and those that are odd under horizontal reflections $\sigma_h\colon z \mapsto -z$ with respect to the V-kagome plane (Fig.~\ref{fig:crystal_structure}(a)). The motivation for this separation is that microscopic theories of LC order find that they are naturally formed of those orbitals that are odd under $\sigma_h$ \cite{li_origin_2023, li_intertwined_2024}. In Fig.~\ref{fig:fermi_surface}, we show the resulting Fermi surface of the full tight-binding model (a) and of the mirror-odd subsector (b).

\noindent\textbf{Classification of loop-current states:} The LC patterns we classify according to two distinct properties: their symmetry transformations under the crystallographic space group and their orbital compositions (i.e., between which types of orbitals do the LCs flow).
Broken symmetry states can be classified according to the irreducible representations (irreps) of the space group of the parent disordered state. While various Ginzburg-Landau theories \cite{denner_analysis_2021, christensen_loop_2022, christensen_theory_2021, holbaek_interplay_2025}, as well as more microscopic symmetry analyses \cite{feng_low-energy_2021, wagner_phenomenology_2023}, have been carried out for charge and LC order in kagome systems, we proceed with an approach that, in addition, takes into account the $3d$ V and planar $5p$ Sb orbital structure. This allows us to explicitly construct the couplings between the LC patterns and the electronic structure described in the previous section. We consider all possible LC patterns that can exist within a $2\times 2$ unit cell and that break translation invariance with ordering vectors $\mathbf{Q}_{\ell=1,2,3}$ which connect the Van Hove points (Fig.~\ref{fig:LCO_patterns}). Regarding the orbital compositions, there are two different categories we consider: LC patterns that impact the phase of nearest-neighbor hopping parameters between V sites, and LC patterns that impact the phase of the hopping parameters between V sites and planar Sb sites. In both of these scenarios, only the 13 orbitals which are odd under $\sigma_h$ are involved. We will later show that the orbital composition is what selects between $d+id$ vs.\ $s^\pm$ pairing.

Using standard group theory techniques \cite{hermele_properties_2008,christensen_loop_2022, wagner_phenomenology_2023}, we classify these patterns according to the irreducible representations of the reduced space group of this extended $2\times 2$ unit cell. The character table and group operations of this reduced space group are detailed in the Supplementary, Sec.~1. These LC patterns belong to three-dimensional irreps, with each pattern in the irrep possessing one of the three ordering vectors $\mathbf{Q}_{\ell}$, which transform into each other under a three-fold rotation. The analysis then yields three fermionic bilinears ($\ell = 1,2,3$)
\begin{equation}
\hat{J}^\ell_{\mathbf{q}} = \sum_{\mathbf{k},\sigma} \hat{c}^\dagger_{\mathbf{k}\sigma} J^\ell(\mathbf{k},\mathbf{k}+\mathbf{q}) \hat{c}_{\mathbf{k}+\mathbf{q}\sigma} \label{eq:current_operator}
\end{equation}
whose expectation values determine the three-component order parameter of the LC state. Here, $\hat{c}_{\mathbf{k}\sigma}$ is a $13$-component spinor that annihilates an electron of spin $\sigma$ and momentum $\mathbf{k}$ in one of the $13$ orbitals that is odd under the mirror symmetry $\sigma_h$. The explicit forms of the $13\times13$ matrices $J^\ell(\mathbf{k},\mathbf{k}+\mathbf{q})$, which are determined by each specific LC pattern, are discussed in the Methods section. We remark that the group theoretic analysis does not fully constrain the loop current patterns, as they must also satisfy local charge conservation at every site \cite{palle_superconductivity_2024,palle_unconventional_2024-1} (Kirchhoff's law). The specific weights of current operators are shown in Table~\ref{tab:lco_pattern_coeffs}, and respect this local conservation law.

More important than the symmetry classification turns out to be the analysis of the current flow within each irrep. In particular, there are several LC states of identical transformation behavior (i.e., that transform under the same irrep), yet with very different microscopic pathways for the coherent circulation of charge. This is shown in Fig.~\ref{fig:LCO_patterns} for LCs that transform under the irreducible representation $mM_{2}^+$ (the $m$ stands for time-reversal odd and the superscript $\pm$ for parity even/odd), which corresponds to Figs.~\ref{fig:LCO_patterns}(a),(b),(c). Notice, in Fig.~\ref{fig:LCO_patterns} we only show LC patterns with ordering wave vector $\mathbf{Q}_3 = (0,\frac{2\pi}{\sqrt{3}})$. For each panel, there are two additional patterns (shown in Supplementary Sec.~2) that are obtained by performing a six-fold rotation around the $z$-axis and which together comprise a three-dimensional space group irrep.
As we shall see, the relative weights of the microscopic pathways depicted in Figs.~\ref{fig:LCO_patterns}(a),(b),(c) -- which are all of same symmetry -- dictate the resulting SC state. Since no experimental constraint exists thus far that favors one pathway over the other, our phase diagrams will show the pairing state as function of the relative weight between these LC pathways.
\noindent\textbf{Effective electron-electron interaction and pairing instabilities}: The interplay between LC order and superconductivity can be analyzed from various points of view. Previous studies have primarily focused on competing LC and SC instabilities \cite{yang_intertwining_2023, park_electronic_2021} or simpler toy models \cite{romer_superconductivity_2022, wu_nature_2021, li_loop-current_2025}. Here, we study the superconductivity mediated by LC fluctuations from the disordered side of the phase diagram, i.e., before the collective LC modes have condensed (Fig.~\ref{fig:crystal_structure}(c)).
This is analogous to studies of SC mediated through the exchange of a nematic, spin magnetic, phononic, or other collective mode \cite{palle_superconductivity_2024,palle_unconventional_2024-1}. 

The effective low-energy electron-electron interaction that is mediated by fluctuating LCs is given by
\begin{equation}
\hat{H}_{{\rm int}} = -\frac{g^2}{2N}\sum_{\mathbf{q}}\sum_{\ell,\ell'=1}^3 [\mathcal{D}_{\text{LC}}(\mathbf{q})]_{\ell\ell'} \hat{J}^{\ell}_{-\mathbf{q}} \hat{J}^{\ell'}_{\mathbf{q}},\label{eq:interaction_hamiltonian}
\end{equation}
where $N$ is the number of unit cells and $g$ is a coupling constant. We approach the modeling of the LC propagator $\mathcal{D}_{\text{LC}}(\mathbf{q})$ in two ways. In the first, we deduce a simple phenomenological form from symmetries and physical considerations.
In the second, we use the random phase approximation (RPA) to calculate the propagator.
As it will turn out, both predict the same pairing symmetries and qualitative behavior for the LC-mediated superconductivity.

To describe $M$-point LCs, the diagonal components of the LC propagator $[\mathcal{D}_{\text{LC}}(\mathbf{q})]_{\ell\ell}$ should be peaked at $\mathbf{Q}_\ell \cong \mathbf{M}_\ell$.
Moreover, since we are approaching the problem from the disordered side, the LC propagator must obey the symmetries of the kagome lattice.
The simplest phenomenological ansatz consistent with these requirements is the following:
\begin{align}
[\mathcal{D}_{\text{LC}}(\mathbf{q})]_{\ell\ell'} = \delta_{\ell\ell'} \big[r + (1-r)f(\mathbf{q} - \mathbf{M}_\ell)\big]^{-1}, \label{eq:phenom_propagator}
\end{align} 
where
$f(\mathbf{q}) = \frac{2}{3} - \frac{2}{9}\big(\cos(\mathbf{q}\cdot\mathbf{a}_1) + \cos(\mathbf{q}\cdot\mathbf{a}_2) + \cos(\mathbf{q}\cdot\mathbf{a}_3)\big)$ with $\mathbf{a}_1 = (1,0)$, $\mathbf{a}_2 = (\frac{1}{2},\frac{\sqrt{3}}{2})$, and $\mathbf{a}_3 = \mathbf{a}_2 - \mathbf{a}_1$. Alternatively, we can explicitly compute the propagator of the loop currents within RPA, in which case it is given by
\begin{align}
&[\mathcal{D}^{-1}_{\text{RPA}}(\mathbf{q},iq_0)]_{\ell\ell'} = \delta_{\ell\ell'} \label{eq:RPA_propagator} \\
&+ \frac{1}{\beta NV}\sum_{k} \text{tr}[\mathcal{G}_0(k) J^\ell(\mathbf{k},\mathbf{k}+\mathbf{q})\mathcal{G}_0(k+q) J^{\ell'}(\mathbf{k}+\mathbf{q},\mathbf{k})]. \nonumber
\end{align}
Here, $\mathcal{G}_0(k)$ is the ($13 \times 13$ matrix) Green function of the tight-binding model, and $J^\ell$ are the $13\times 13$ matrices from Eq.~\eqref{eq:current_operator}. Due to imperfect nesting of the band structure, the RPA propagator will generally be peaked away from the $M$-points, and has structure elsewhere in the Brillouin zone which does not change the pairing state. We include a plot in the Supplementary Sec.~3 comparing the RPA propagator in Eq.~\eqref{eq:RPA_propagator} with the phenomenological one from Eq.~\eqref{eq:phenom_propagator}.

This interaction can be projected onto the Cooper channel to give a gap equation. Since we are interested in the leading instability, at weak-coupling we may linearize this gap equation to obtain an eigenvalue problem:
\begin{equation}
\lambda\Delta_{n}(\mathbf{k}) = - \sum_{n'}\oint_{\text{FS}_{n'}}  \frac{U^{s/t}_{nn'}(\mathbf{k},\mathbf{k}')\Delta_{n'}(\mathbf{k}')\text{d}{k}'}{|\nabla_{\mathbf{k}'}\xi_{\mathbf{k}' n'}|(2\pi)^2}. \label{eq:gap_equation}
\end{equation}
Here, $\Delta_{n}(\mathbf{k})$ is the singlet/triplet pairing amplitude in band $n$ with wave vector $\mathbf{k}$, $U^{s/t}_{nn'}(\mathbf{k},\mathbf{k}')$ is the interaction in the singlet/triplet Cooper channel between points $\mathbf{k},\mathbf{k}'$ in bands $n,n'$, respectively, and $\xi_{\mathbf{k} n}$ is the dispersion.
The singlet and triplet channels are decoupled because the band Hamiltonian is inversion-symmetric. Since neither LC fluctuations nor the band Hamiltonian (which has no spin-orbit coupling) break the spin rotation symmetry, the triplet channel $\mathbf{d}$-vector may point in any direction, $\mathbf{d}_n(\mathbf{k}) = \Delta_{n}(\mathbf{k}) \hat{\mathbf{n}}$.
If all $\lambda < 0$, then there is no SC instability. We therefore search for the largest $\lambda > 0$ which corresponds to a SC instability with the largest $T_c \propto e^{-1/\lambda}$.
When $U(\mathbf{k},\mathbf{k}') < 0$ is attractive, then this cancels the leading minus sign in the gap equation and $\Delta$ generically has the same sign at $\mathbf{k}$ and $\mathbf{k}'$. Conversely, if $U(\mathbf{k},\mathbf{k}') > 0$ is repulsive, then the $\lambda > 0$ solutions favor a situation wherein $\Delta(\mathbf{k})$ and $\Delta(\mathbf{k}')$ have opposite signs.
\begin{figure*}[t]
\centering
\includegraphics[width=\textwidth]{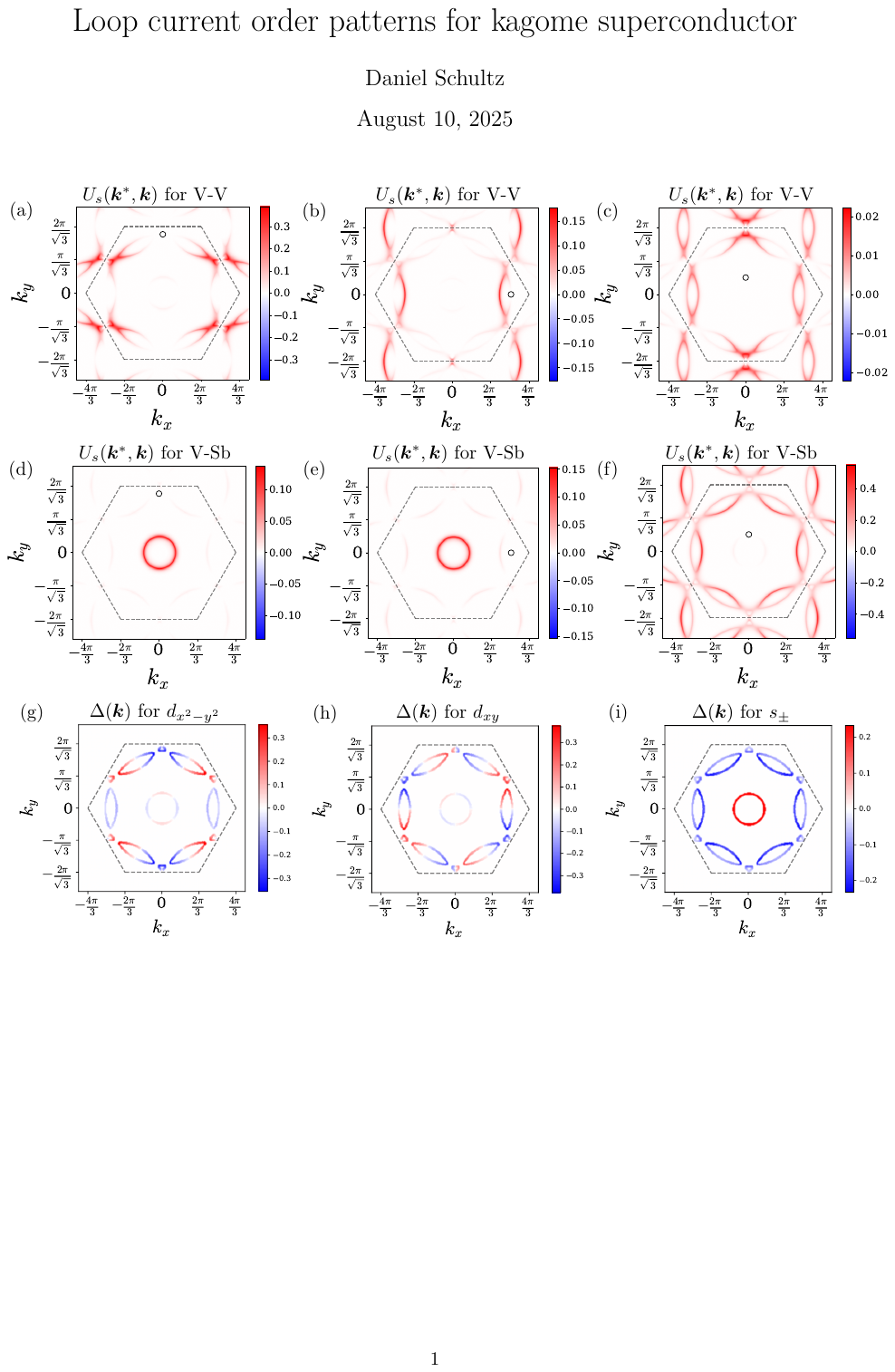}
\caption{\textit{Projected pairing kernels and gap functions.} The singlet-channel pairing interaction for two LC patterns (a)--(f) and examples of SC gap functions they result in (g)--(i).
Under (a)--(f), one momentum of the interaction $\mathbf{k}^*$ is fixed to the white point, which is on a Fermi surface, while the other momentum $\mathbf{k}$ is varied across the remaining Fermi surface area, with red color indicating the degree of repulsion.
As mentioned in the discussion, the singlet interactions is always repulsive for interactions mediated by LCs.
Panels (a)--(c) show the interaction for the V-V $mM_{2}^+$ pattern of Fig.~\ref{fig:LCO_patterns}(a).
Panels (d)--(f) correspond to the V-Sb $mM_{2}^+$ patterns of Fig.~\ref{fig:LCO_patterns}(c).
In (g) we show the basis gap function for $d_{x^2-y^2}$ pairing, in (h) for $d_{xy}$, and in (i) for $s^{\pm}$. In hexagonal systems, $d_{x^2-y^2}$ and $d_{xy}$ belong to the same two-dimensional irrep, namely $E_{2g}$. All plots in this figure were generated for mass $r = 0.3$, and used the phenomenological propagator from Eq.~\eqref{eq:phenom_propagator}. Using the RPA propagator yields indistinguishable results for the pairing states.} \label{fig:interactions_and_gaps}
\end{figure*}
As discussed in related contexts \cite{lederer_enhancement_2015,palle_superconductivity_2024}, under certain conditions $\lambda_{\text{max}}$ can diverge upon approaching a quantum-critical point. While our weak-coupling approach clearly breaks down in this regime, it is  established that strong-coupling treatments can regularize this divergence, yielding the maximum in $T_c$ at the transition, as sketched in Fig.~\ref{fig:crystal_structure}(c), without altering the pairing symmetry identified in the weak-coupling analysis \cite{son_superconductivity_1999, abanov_coherent_2001, abanov_quantum-critical_2001, moon_quantum-critical_2010, esterlis_quantum_2025}. There is therefore good reason to believe that our weak-coupling theory -- which enables us to address a problem involving a complex pairing interaction with a large number of orbitals per unit cell -- can identify the leading pairing symmetry. We also note that the charge bond ordered (with potential loop current component) state in the kagome superconductors experiences a first-order transition\cite{zheng_emergent_2022} to a disordered state (at around 2 GPa in \ce{CsV3Sb5}\cite{stier_pressure-dependent_2024}), meaning that the mass never truly becomes zero at the transition.

For LC patterns that only have currents flowing between vanadium atoms (patterns (a),(b),(d) in Fig.~\ref{fig:LCO_patterns}), the dominant pairing instability (at the level of the linearized gap equation) is in the $E_{2g} = \{d_{x^2-y^2},d_{xy}\}$ irrep. To see why this is the case, in Figs.~\ref{fig:interactions_and_gaps}(a)--(c) we have plotted the Fermi-surface-projected singlet-channel pairing interaction $U_{s}(\mathbf{k},\mathbf{k}')$, which is the same interaction appearing in Eq.~\eqref{eq:gap_equation}, for the V-V LC pattern of Fig.~\ref{fig:LCO_patterns}(a). 
As one can see in Fig.~\ref{fig:interactions_and_gaps}(a), the interaction between the white point near $\mathbf{M}_3$ and the other $\mathbf{M}$-points is repulsive. The best way to achieve pairing from this repulsive interaction is through $d_{x^2-y^2}$ or $d_{xy}$ pairing since the gap function changes sign between the $\mathbf{M}$-points. The corresponding SC gap functions are displayed in Figs.~\ref{fig:interactions_and_gaps}(g),(h). After introducing non-linear corrections to the linearized gap equation, we find that the time-reversal symmetry-breaking state $d_{x^2-y^2} + id_{xy}$ is favored. The $d_{x^2-y^2}+id_{xy}$ state is fully gapped, in contrast to $d_{x^2-y^2}$ or $d_{xy}$, each of which has nodal lines.
Importantly, because there is almost no coupling between the outer sheets of the Fermi surface and the planar Sb pocket near the $\mathbf{\Gamma}$-point, the SC gap function can be negligibly small on the inner $\mathbf{\Gamma}$-pocket, with a sign that has no reason to be opposite to the outer Fermi surface sheets.
In turn, this implies that $d_{x^2-y^2} + id_{xy}$, or more simply $d+id$, pairing is insensitive to the presence of the $\mathbf{\Gamma}$-pocket.
The triplet-channel interactions are, for completeness, shown in the Supplementary, Sec.~4.

On the other hand, if we have currents flowing between V and Sb atoms (the pattern in Fig.~\ref{fig:LCO_patterns}(c)), there exists a strong repulsion between the circular Fermi surface around $\mathbf{\Gamma}$ and the parts of the Fermi surface near the $\mathbf{M}$-points, as shown in Figs.~\ref{fig:interactions_and_gaps}(d)--(f). This indicates that the gap on the outer sheets of the Fermi surface must have opposite sign compared to the Sb sheet near $\mathbf{\Gamma}$. Consequently, the leading solution for this case has $s^{\pm}$ symmetry, as indicated by the gap structure in Fig.~\ref{fig:interactions_and_gaps}(i). The pattern of Fig.~\ref{fig:LCO_patterns}(e) has both V-V and V-Sb loop currents coexisting, so whether $d+id$ or $s^{\pm}$ is favored depends on the specific ratio of energy scales between V-V interactions and V-Sb interactions. In the case we have chosen, this pattern yields $s^{\pm}$ pairing. The fact that fluctuating V-V LC patterns favor $d+id$ pairing, while fluctuating V-Sb LC patterns favor $s^{\pm}$ pairing, is the primary result of this paper.

\begin{figure}[t!]
\centering
\includegraphics[width=\columnwidth]{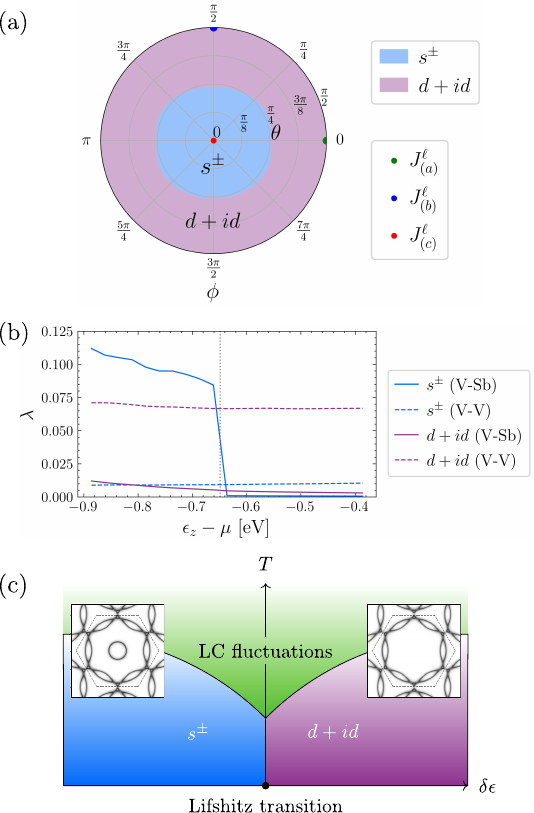}
\caption{\textit{Phase diagram of pairing symmetries under various dominant LC pathway.} (a) A polar plot showing how the dominant pairing instability depends on the LC pattern, as parametrized by $\hat{J}^{\ell} = \sin\theta\cos\phi \hat{J}^{\ell}_{(a)} + \sin\theta\sin\phi \hat{J}^{\ell}_{(b)} + \cos\theta \hat{J}^{\ell}_{(c)}$. The variables $\hat{J}^\ell_{(a)},\hat{J}^\ell_{(b)},\hat{J}^\ell_{(c)}$ refer to the current patterns of Figs.~\ref{fig:LCO_patterns}(a),(b),(c), respectively. The plot is periodic in $\phi$. $\theta \in [\frac{\pi}{2}, \pi]$ is not shown because $\hat{J}^{\ell}$ and $-\hat{J}^{\ell}$ both give the same interaction. When $\theta$ (radial variable) is small, the currents flow predominantly between V-Sb, driving $s^{\pm}$ pairing.
(b) The phase diagram as a function of $\epsilon_z$, which is the tuning parameter for the $\mathbf{\Gamma}$-point pocket, for $\hat{J}_{(a)}$ (dashed lines) and $\hat{J}_{(c)}$ (solid lines) exchange.
Since $\hat{J}_{(a)}$ is a current pattern flowing only between V-V, the $\mathbf{\Gamma}$-pocket plays essentially no role, so its removal (vertical dotted line) does not affect the leading pairing eigenvalue.
In contrast, because $\hat{J}_{(c)}$ flows only between V-Sb, which strongly couples the $\mathbf{\Gamma}$-pocket and outer Fermi surface, the removal of the $\mathbf{\Gamma}$-pocket destroys the $s^{\pm}$ superconductivity; only weak interactions between the $\mathbf{M}$-points on the outer Fermi surface remain, yielding $d+id$ pairing. (c) A schematic phase diagram illustrating the quantitative result of (b), wherein the removal of the $\mathbf{\Gamma}$ pocket at the Lifshitz transition destroys the $s^\pm$ state. Panels (a) and (b) were calculated using the RPA LC propagator from Eq.~\eqref{eq:RPA_propagator}.} \label{fig:phase_diagrams}
\end{figure}

Having studied the two cases separately, we now address what happens if we consider a generic superposition of V-V and V-Sb LC patterns. In Fig.~\ref{fig:LCO_patterns}, patterns (a),(b),(c) all belong to the same irrep $mM_{2}^+$ of the reduced space group. This means that they can be added together with arbitrary coefficients to form a new current pattern belonging to the same irrep. We therefore study how the leading SC state depends on the normalized combination $\hat{J}^{\ell} = \sin\theta\cos\phi \hat{J}^{\ell}_{(a)} + \sin\theta\sin\phi \hat{J}^{\ell}_{(b)} + \cos\theta \hat{J}^{\ell}_{(c)}$.
The subscripts (a),(b),(c) refer to the patterns in Figs.~\ref{fig:LCO_patterns}(a),(b),(c). The corresponding phase diagram is illustrated in Fig.~\ref{fig:phase_diagrams}(a). Evidently, when $\theta$ is small, we have predominantly $s^{\pm}$ pairing, as expected given that the current pattern is predominantly V-Sb of the type shown in Fig.~\ref{fig:LCO_patterns}(c), with interactions depicted in Figs.~\ref{fig:interactions_and_gaps}(d)--(f). If, instead, $\theta$ is close to $\frac{\pi}{2}$, then the LC pattern is dominated by currents flowing between V-V sites, and consequently the interaction closely resembles those of Figs.~\ref{fig:interactions_and_gaps}(a)--(c). This leads to $d+id$ pairing, as expected. We include in the Supplementary Sec.~4 the plots of the projected pairing interaction for all loop current patterns, in addition to the ones shown in Fig.~\ref{fig:interactions_and_gaps}.
The next issue to address is the fate of the SC solution arising from the V-Sb LC pattern (Fig.~\ref{fig:LCO_patterns}(c)) when the $\mathbf{\Gamma}$-pocket is removed. Experimentally, in \ce{CsV3Sb5} the $\mathbf{\Gamma}$-pocket undergoes a Lifshitz transition (i.e., it moves above the Fermi level) at approximately $7.5$ GPa, which is around the same point at which the superconductivity disappears \cite{zhang_pressure-induced_2021}. In our tight-binding model, we can tune the on-site potential $\epsilon_{z}$ of the planar Sb $p_z$ orbital to move this circular pocket above the Fermi energy. Upon doing this, we find that the $s^{\pm}$ superconductivity is strongly suppressed at the Lifshitz transition (signaled by the vertical dotted line in Fig.~\ref{fig:phase_diagrams}(b)), and that the dominant eigenvalue once again falls into the $E_{2g}$ irrep. The leading eigenvalue and its symmetry classification as a function of $\epsilon_{z}$ is depicted in Fig.~\ref{fig:phase_diagrams}(b).

\noindent{\large\textbf{Discussion}}

\noindent In our analysis of superconductivity due to LC fluctuations, we have considered a wide array of different LC patterns with translation symmetry-breaking ($2\times 2$ unit cell) in line with scanning tunneling microscopy and X-ray scattering experiments. We considered states of even and odd parity, with and without currents flowing to the planar Sb atoms. The main result of our work is that current patterns flowing only between V yield $d+id$ SC, yet current patterns also flowing to planar Sb yield $s^{\pm}$ SC, regardless of the parity of the LC state. These results can be understood as follows.

It is a general feature of boson-mediated pairing that a time-reversal even boson mediates an attractive (in the singlet Cooper channel) interaction, whereas a time-reversal odd boson mediates a repulsive interaction \cite{palle_unconventional_2024-1}. Common examples of the former are attraction mediated by phonons or nematic fluctuations. On the other hand, spin fluctuations, and the very LC fluctuations described in this paper, mediate repulsive interactions. It should therefore come as no surprise that the singlet-channel interactions in Fig.~\ref{fig:interactions_and_gaps}(a)--(f) are nonnegative (i.e., repulsive) everywhere in the Brillouin zone. As long as the repulsive interaction is between distinct points on the Fermi surface (i.e., for finite momentum exchange), unconventional SC solutions are typical \cite{maiti_superconductivity_2013}. The fact that the LC propagator peaks at (or nearby) $\mathbf{M}$-points means that the repulsion is large between the different $\mathbf{M}$-points, highlighting the effectiveness of translation symmetry-breaking loop currents in driving superconductivity.
Furthermore, the pairing eigenvalue $\lambda$ is enhanced as the boson mass $r \to 0$, making the interaction most efficient near a LC quantum-critical point.

The V-V patterns, regardless of whether they have even or odd parity, consist of electrons on different V sublattices interacting with (and, furthermore, repelling) one another. In momentum space, such an interaction couples electrons at different $\mathbf{M}$-points. This incentivizes sign-changing of the gap function between different $\mathbf{M}$-points, and is the origin of the $d_{x^2-y^2} + id_{xy}$ pairing. On the other hand, in a V-Sb pattern, electrons from V sites repel the electrons from the in-plane Sb. In momentum space, this leads to a strong repulsion between electrons near $\mathbf{M}$-points and the circular Fermi surface around the $\mathbf{\Gamma}$-point. Such a strong repulsion makes the gap function change its sign between the inner $\mathbf{\Gamma}$-pocket and the $\mathbf{M}$-point electrons, giving $s^\pm$ pairing. It is interesting to note that pairing states of different symmetries are favored by LC patterns with the same symmetry, promoted by different orbital compositions. This is reminiscent of the physics of iron-based superconductors, for which spin fluctuations can favor either $s^\pm$ or $d_{x^2-y^2}$ wave pairing \cite{fernandes_iron_2022}.

How does our theory of LC-induced unconventional pairing compare with experimental observations of the superconducting gap structure in \ce{AV3Sb5}? Both candidate pairing states -- $s^\pm$ and chiral $d+id$ -- are fully gapped (apart from possible accidental nodes), consistent with experimental observations \cite{zhong_nodeless_2023, deng_evidence_2024, zhang_nodeless_2023}. Moreover, recent quasiparticle interference measurements \cite{deng_chiral_2024} indicate a large SC gap on the $\mathbf{\Gamma}$-pocket Fermi surface, which favors the $s^{\pm}$ pairing candidate.
The disappearance of superconductivity in \ce{CsV3Sb5} around $7.5$ GPa \cite{zhang_pressure-induced_2021, chen_highly_2021}, coinciding with the Lifshitz transition that eliminates the $\mathbf{\Gamma}$ pocket, further underscores the importance of planar Sb states in mediating superconductivity. Remarkably, a second SC dome appears at approximately $16.5$ GPa \cite{zhang_pressure-induced_2021, chen_highly_2021}, which our theory naturally interprets as a transition from the $s^\pm$ state to the $d+id$ state, as shown in Fig.~\ref{fig:phase_diagrams}(c).
Superconductivity near ambient pressure being of the $s^{\pm}$ type is also consistent with electron irradiation experiments \cite{roppongi_bulk_2023}, which show that the SC state remains robust in the presence of charge impurities. This is expected, as $s^{\pm}$ pairing is known to be more resilient to impurity scattering than chiral $d+id$ pairing \cite{hoyer_pair_2015}.
Finally, $\mu$SR experiments at low pressure ($P < 7.5$ GPa) observe a change in the muon relaxation rate below $T_c$, including in systems lacking CDW order \cite{guguchia_tunable_2023,deng_evidence_2024}. This has been interpreted as evidence for time-reversal symmetry-breaking, which would na\"ively suggest $d+id$ pairing in this regime. However, recent insights from the case of \ce{Sr2RuO4} show that similar $\mu$SR signatures can arise from closely competing pairing states, with time-reversal symmetry being locally broken near strain inhomogeneities and dislocations \cite{willa_inhomogeneous_2021, yuan_strain-induced_2021, andersen_spontaneous_2024}. Our theory provides a natural explanation for such near-degenerate superconducting states. We therefore expect that local $E_{2g}$ strain fields may stabilize time-reversal symmetry-breaking admixtures of the dominant $s^{\pm}$ and subleading $d+id$ pairing components. Overall, the chiral $d+id$ state remains a viable candidate in parts of the phase diagram (Fig.~\ref{fig:phase_diagrams}). The pressure-induced transition between $s^{\pm}$ and $d+id$ pairing states comes naturally as an explanation for the observations of Refs.~\cite{zhang_pressure-induced_2021, chen_highly_2021} within our framework.

One aspect that we have neglected is the large hexagonal-shaped Fermi surface that passes close to the $\mathbf{M}$-points and that is due to orbitals which are even under $\sigma_h\colon z \mapsto -z$. This part of the Fermi surface may be identified in Fig.~\ref{fig:fermi_surface}(a) (but is, of course, absent in Fig.~\ref{fig:fermi_surface}(b)). To understand the impact of these degrees of freedom, we observe that, because our interactions are generated purely through fluctuating LC modes made up of $\sigma_h$-odd LC patterns, they only couple $\sigma_h$-odd orbitals to other $\sigma_h$-odd orbitals. However, even at $k_z = 0$, the $\sigma_h$-even orbitals should couple to $\sigma_h$-odd orbitals via additional interactions, such as the Coulomb interaction. The gap opened up by either $d+id$ or $s^{\pm}$ superconductivity on the $\sigma_h$-odd bands is therefore expected to induce a gap on the $\sigma_h$-even bands as well.

In conclusion, fluctuating translation symmetry-breaking LC patterns are an effective pairing glue for superconductivity. The two candidate states we find are $d+id$ and $s^{\pm}$, both of which are fully gapped states. In particular, to obtain $s^{\pm}$ superconductivity, little explored LC patterns which traverse between V and Sb atoms must be included in the analysis. Our results for the $s^{\pm}$ state are consistent with the crucial role of planar Sb states \cite{bhandari_first-principles_2024, bhandari_pressure_2025} for superconductivity below $7.5$ GPa \cite{chen_highly_2021, zhang_pressure-induced_2021}, whereas our results for the $d+id$ state also present a potential pairing mechanism for superconductivity beyond $16.5$ GPa \cite{chen_highly_2021, zhang_pressure-induced_2021}, at which point no $\mathbf{\Gamma}$-pocket remains \cite{bhandari_first-principles_2024, bhandari_pressure_2025}. The pairing symmetry of the superconductivity depends crucially on the microscopic pathway of the loop current, highlighting the significance of orbital-selective LC interactions, and their implications for couplings on the Fermi surface.

\vspace{6pt}
\noindent{\large\textbf{Methods}}

\begin{figure}[t!]
\centering
\includegraphics[width=0.95\columnwidth]{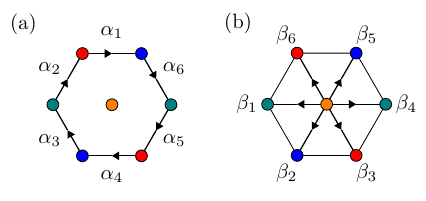}
\caption{\textit{Labels for coefficients of the current operator within a single hexagon.} In both cases, we take the convention that setting the coefficient positive yields the indicated direction of current. (a) The case of V-V currents. (b) The case of V-Sb currents.} \label{fig:current_bond_labels}
\end{figure}

\noindent\textbf{Effective electron-electron interaction}: Here we present the explicit expressions for the coupling between the electrons and the LC boson corresponding to the several LC patterns drawn in Fig.~\ref{fig:LCO_patterns} of the main text. Consider a hexagon as shown in Fig.~\ref{fig:current_bond_labels}. The index $n$ labels the unit cell that the planar Sb site lives in. The LC operator on a single hexagon has two different forms for the two cases of V-V or V-Sb patterns, respectively:
\begin{align}
\hat{J}_n ={}& \sum_{c = 1}^6 \alpha_c \hat{j}_c, \qquad \text{V-V pattern,} \\
\hat{J}_n ={}& \sum_{c = 1}^6 \beta_c \hat{j}_c, \qquad \text{V-Sb pattern.}
\end{align}
The meanings of the coefficients $\alpha_c$ and $\beta_c$ are indicated in Fig.~\ref{fig:current_bond_labels}. The subscript $c$ labels the different bonds and $\hat{j}_c$ is the current operator on the specified bond; see Fig.~\ref{fig:current_bond_labels}. By specifying the $\alpha_c,\beta_c$ coefficients, one can construct the current operator $\hat{J}_n$ and hence $\hat{J}_{\mathbf{q}}$. The coefficients for these patterns are listed in Table~\ref{tab:lco_pattern_coeffs}.
By calculating the Fourier transform $\hat{J}_{\mathbf{q}} = \sum_n e^{-i\mathbf{R}_n\cdot\mathbf{q}} \hat{J}_n$, these couplings lead to the following interaction term
\begin{align}
S_{\text{LC-el}} = -\frac{g}{\beta N} \sum_{iq_0,\mathbf{q}} \sum_{\ell=1}^3 J^\ell_{q} \Phi^\ell_{q},
\end{align}
where the sum over $\ell$ goes over the three different $\mathbf{M}$-points.
Here, the fermionic bilinears couple to real, time-reversal odd bosons which we call $\Phi$, representing the collective LC modes. There are three bosons, one for each $\mathbf{M}$-point. The kinetic term for the bosons consists of the inverse propagator:
\begin{equation}
S_{\Phi} = \frac{1}{2\beta N}\sum_{q} \sum_{\ell,\ell'=1}^3 [\mathcal{D}^{-1}_{\text{LC}}(\mathbf{q})]_{\ell\ell'} \Phi^\ell_{-q} \Phi^{\ell'}_{q},
\end{equation}
The loop current propagator $[\mathcal{D}_{\text{LC}}(\mathbf{q})]_{\ell\ell'}$ has its maximum at the ordering wave vector of the boson.

\begin{table*}[t!]
\centering
{\renewcommand{\arraystretch}{1.4}
\renewcommand{\tabcolsep}{13pt}
\begin{tabular}{cccc}
\hline\hline
Pattern & Wave vector(s) & $(\alpha_1,\alpha_2,\alpha_3,\alpha_4,\alpha_5,\alpha_6)$ & $(\beta_1,\beta_2,\beta_3,\beta_4,\beta_5,\beta_6)$ \\ \hline
(a) & $\mathbf{M}_1,\mathbf{M}_2,\mathbf{M}_3$ & $\frac{1}{\sqrt{6}}(1,1,1,1,1,1)$ & $(0,0,0,0,0,0)$ \\ \hline
 & $\mathbf{M}_1$ & $\frac{1}{\sqrt{12}}(1,1,-2,1,1,-2)$ & $(0,0,0,0,0,0)$ \\
(b) & $\mathbf{M}_2$ & $\frac{1}{\sqrt{12}}(1,-2,1,1,-2,1)$ & $(0,0,0,0,0,0)$ \\
 & $\mathbf{M}_3$ & $\frac{1}{\sqrt{12}}(-2,1,1,-2,1,1)$ & $(0,0,0,0,0,0)$ \\ \hline
 & $\mathbf{M}_1$ & $(0,0,0,0,0,0)$ & $\frac{1}{2}(-1,1,0,-1,1,0)$ \\
(c) & $\mathbf{M}_2$ & $(0,0,0,0,0,0)$ & $\frac{1}{2}(1,0,-1,1,0,-1)$ \\
 & $\mathbf{M}_3$ & $(0,0,0,0,0,0)$ & $\frac{1}{2}(0,-1,1,0,-1,1)$ \\ \hline
 & $\mathbf{M}_1$ & $(0.358, -0.358, -0.494, -0.358, 0.358, 0.494)$ & $(0,0,0,0,0,0)$ \\
(d) & $\mathbf{M}_2$ & $(-0.358, -0.494, -0.358, 0.358, 0.494, 0.358)$ & $(0,0,0,0,0,0)$ \\
 & $\mathbf{M}_3$ & $(0.494, 0.358, -0.358, -0.494, -0.358,0.358)$ & $(0,0,0,0,0,0)$ \\ \hline
 & $\mathbf{M}_1$ & $-\frac{0.77535}{\sqrt{6}}(1, -1, 1, -1, 1 -1)$ & $\frac{0.63153}{\sqrt{6}}(-1,1,1,1,-1,-1)$ \\
(e) & $\mathbf{M}_2$ & $-\frac{0.77535}{\sqrt{6}}(-1, 1, -1, 1, -1, 1)$ & $\frac{0.63153}{\sqrt{6}}(1,1,1,-1,-1,-1)$ \\
 & $\mathbf{M}_3$ & $-\frac{0.77535}{\sqrt{6}}(-1, 1, -1, 1, -1, 1$ & $\frac{0.63153}{\sqrt{6}}(-1,-1,1,1,1,-1)$ \\
\hline\hline
\end{tabular}}
\caption{Current patterns within a unit cell. The pattern labels reference the patterns in Fig.~\ref{fig:LCO_patterns} of the main text. In the main text, only the patterns with ordering wave vector $\mathbf{Q}_3$ are shown. The cases (a),(b),(c) have very simple coefficients because the Kirchhoff law is enforced for symmetry reasons. In contrast, the cases (d),(e) require fine-tuning of the parameters.} \label{tab:lco_pattern_coeffs}
\end{table*}

\noindent\textbf{Integrating out the LC boson}: We integrate out the LC boson by performing the following Gaussian integral:
\begin{gather}
\begin{gathered}
\int \exp\left(-\frac{1}{\beta N}\sum_{q}\left[\frac{1}{2} [\mathcal{D}^{-1}_{\text{LC}}(\mathbf{q})] |\Phi_{q}|^2 - g\Phi_{q} J_{-q}\right]\right) \mathcal{D}\Phi = \\
= \exp\left(\frac{g^2}{2\beta N}\sum_{q} \mathcal{D}_{\text{LC}}(\mathbf{q}) J_{-q} J_{q}\right).
\end{gathered}
\end{gather}
\begin{widetext}
This generates the following interaction which is mediated by LC fluctuations:
\begin{align}
\begin{aligned}
\hat{H}_{\text{int}} ={}& -\frac{g^2}{2N}\sum_{\mathbf{q}}\sum_{\ell,\ell'=1}^3 [\mathcal{D}_{\text{LC}}(\mathbf{q})]_{\ell\ell'} :\hat{J}^\ell_{-\mathbf{q}} \hat{J}^{\ell'}_{\mathbf{q}}: \\
={}& \frac{1}{2N}\sum_{\substack{\mathbf{k}_1\mathbf{k}_2\mathbf{k}_2'\mathbf{k}_1' \\ abb'a' \\ \sigma_1\sigma_1'\sigma_2'\sigma_2} } V^{\sigma_1\sigma_2\sigma_2'\sigma_1'}_{abb'a'}(\mathbf{k}_1,\mathbf{k}_2,\mathbf{k}_2',\mathbf{k}_1') \hat{c}^\dagger_{\mathbf{k}_1 a \sigma_1} \hat{c}^\dagger_{\mathbf{k}_2 b \sigma_2} \hat{c}_{\mathbf{k}_2',b',\sigma_2'} \hat{c}_{\mathbf{k}_1',a',\sigma_1'},
\end{aligned}
\end{align}
whereby now we explicitly indicate the sublattice/orbital indices $a,b,a',b'$ for the fermions. The interaction vertex is then defined according to
\begin{align}
V_{abb'a'}^{\sigma_1\sigma_2\sigma_2'\sigma_1'}(\mathbf{k}_1,\mathbf{k}_2,\mathbf{k}_2',\mathbf{k}_1') = -g^2 \delta_{\mathbf{k}_1+\mathbf{k}_2,\mathbf{k}_1'+\mathbf{k}_2'} \delta_{\sigma_1\sigma_1'}\delta_{\sigma_2\sigma_2'}\sum_{\ell=1}^3 [\mathcal{D}_{\text{LC}}(\mathbf{k}_1-\mathbf{k}_1')]_{\ell\ell'} J^\ell_{aa'}(\mathbf{k}_1,\mathbf{k}_1') J^{\ell'}_{bb'}(\mathbf{k}_2,\mathbf{k}_2').
\end{align}
Next, we explicitly antisymmetrize the interaction (of course, fermion statistics enforces this, but it is nonetheless useful to ensure the matrix elements also obey antisymmetrization)
\begin{align}
U_{abb'a'}^{\sigma_1\sigma_2\sigma_2'\sigma_1'}(\mathbf{k}_1,\mathbf{k}_2,\mathbf{k}_2',\mathbf{k}_1') = V_{abb'a'}^{\sigma_1\sigma_2\sigma_2'\sigma_1'}(\mathbf{k}_1,\mathbf{k}_2,\mathbf{k}_2',\mathbf{k}_1') - V_{bab'a'}^{\sigma_2\sigma_1\sigma_2'\sigma_1'}(\mathbf{k}_2,\mathbf{k}_1,\mathbf{k}_2',\mathbf{k}_1')
\end{align}
so that the interaction is given by
\begin{align}
\hat{H}_{\text{int}} ={}& \frac{1}{4N}\sum_{\substack{\mathbf{k}_1\mathbf{k}_2\mathbf{k}_2'\mathbf{k}_1' \\ abb'a' \\ \sigma_1\sigma_1'\sigma_2'\sigma_2} } U^{\sigma_1\sigma_2\sigma_2'\sigma_1'}_{abb'a'}(\mathbf{k}_1,\mathbf{k}_2,\mathbf{k}_2',\mathbf{k}_1') \hat{c}^\dagger_{\mathbf{k}_1 a \sigma_1} \hat{c}^\dagger_{\mathbf{k}_2 b \sigma_2} \hat{c}_{\mathbf{k}_2',b',\sigma_2'} \hat{c}_{\mathbf{k}_1',a',\sigma_1'}.
\end{align}

\noindent\textbf{Linearized gap equation}: We may now project onto the Cooper channel by requiring that $\mathbf{k} := \mathbf{k}_1=-\mathbf{k}_2$ and $\mathbf{k}' := -\mathbf{k}_2'=\mathbf{k}_1'$. This yields a Cooper channel interaction $U^{\sigma_1\sigma_2\sigma_2'\sigma_1'}_{abb'a'}(\mathbf{k},-\mathbf{k},-\mathbf{k}',\mathbf{k}')$.
The correpsonding BCS-type Hamiltonian is
\begin{align}
\hat{H} ={}& \sum_{\mathbf{k} ab\sigma} \hat{c}^\dagger_{\mathbf{k} a\sigma} [H_0(\mathbf{k})]_{ab} \hat{c}_{\mathbf{k} b\sigma} + \frac{1}{4N}\sum_{\substack{\mathbf{k}\mathbf{k}' abb'a' \\ \sigma_1\sigma_1'\sigma_2'\sigma_2}} U^{\sigma_1\sigma_2\sigma_2'\sigma_1'}_{abb'a'}(\mathbf{k},-\mathbf{k},-\mathbf{k}',\mathbf{k}') \hat{c}^\dagger_{\mathbf{k} a\sigma_1}\hat{c}^\dagger_{-\mathbf{k}, b\sigma_2}\hat{c}_{-\mathbf{k}',b'\sigma_2'} \hat{c}_{\mathbf{k}'a',\sigma_1'}.
\end{align}
By defining the gap function as
\begin{align}
\Delta_{ab\sigma_1\sigma_2}(\mathbf{k}) ={}& -\frac{1}{2N}\sum_{\mathbf{k}' b'a'\sigma_2'\sigma_1'} U^{\sigma_1\sigma_2\sigma_2'\sigma_1'}_{abb'a'}(\mathbf{k},-\mathbf{k},-\mathbf{k}',\mathbf{k}') \langle \hat{c}_{-\mathbf{k}'b'\sigma_2'} \hat{c}_{\mathbf{k}'a'\sigma_1'}  \rangle,
\end{align}
we can perform the standard construction of the linearized gap equation \cite{palle_unconventional_2024-1} to obtain
\begin{align}
\begin{aligned}
\frac{\Delta_{ab\sigma_1\sigma_2}(\mathbf{k})}{\log(\beta \hbar \omega_c 2e^\gamma/\pi)} ={}& -\frac{1}{2}\sum_{nb'a'}  \oint_{\xi_n(\mathbf{k}')=0} U^{\sigma_1\sigma_2\sigma_2'\sigma_1'}_{abb'a'}(\mathbf{k},-\mathbf{k},-\mathbf{k}',\mathbf{k}') [T(-\mathbf{k}')]_{b'n} [T(\mathbf{k}')]_{a'n} \\
&\times \sum_{cd\sigma_2'\sigma_1'} [T^\dagger(\mathbf{k}')]_{nc} [T^\dagger(-\mathbf{k}')]_{nd} [\Delta(\mathbf{k}')]_{cd\sigma_1'\sigma_2'}  \frac{\text{d} \mathbf{k}'}{|\nabla_{\mathbf{k}'}\xi_n(\mathbf{k}')| (2\pi)^2}.
\end{aligned}
\end{align}
Here, the matrix $T$ is the change of basis matrix between orbitals and bands, defined through $[H_0(\mathbf{k})]_{ab} = \sum_n[T(\mathbf{k})]_{an}\xi_n(\mathbf{k})[T^\dagger(\mathbf{k})]_{nb}$. Everything can appropriately be transformed to the band basis (here, $n$ labels bands, whereas $a,b,c,d$ label orbitals). The above linearized gap equation can be recast in the band basis, and then split into singlet and triplet components, yielding the final form of the gap equation found in the main text:
\begin{align}
\lambda \Delta^{s/t}(\mathbf{k}) ={}& - \oint_{\text{FS}} U_{s/t}(\mathbf{k},\mathbf{k}') \Delta^{s/t}(\mathbf{k}') \frac{\text{d} {k}'}{|\nabla\xi(\mathbf{k}')|(2\pi)^2}
\end{align}
Here, $U_{s/t}(\mathbf{k},\mathbf{k}')$ is the interaction in the singlet/triplet channel. It is related to the earlier matrix elements of the interaction in the following way:
\begin{equation}
U^{AB}(\mathbf{k}_m,\mathbf{k}_n) = \sum_{\substack{\sigma_1\sigma_2\sigma_2'\sigma_1' \\ abb'a'}} [\sigma^A\sigma^y]_{\sigma_2\sigma_1} \frac{[T^\dagger(\mathbf{k})]_{m a}[T^\dagger(-\mathbf{k})]_{mb} U^{\sigma_1\sigma_2\sigma_2'\sigma_1'}_{abb'a'}(\mathbf{k},-\mathbf{k},-\mathbf{k}',\mathbf{k}') [T(-\mathbf{k}')]_{b'n} [T(\mathbf{k}')]_{a'n}}{4}[\sigma^y\sigma^B]_{\sigma_1'\sigma_2'}
\end{equation}
This matrix element is proportional to $\delta_{AB}$ because of SU(2) spin invariance, and $m,n$ are actually just set by the point on the Fermi surface since for any given $\mathbf{k}$-point on the Fermi surface, only one band crosses it. Thus $U^{s}$ corresponds to $A = B = 0$, and $U^{t}$ to $A = B = 1,2,3$.
This is the interaction that is plotted in Fig.~\ref{fig:interactions_and_gaps} of the main text.
\end{widetext} 

\bibliographystyle{unsrtnat}


\begin{acknowledgements}
We are grateful to Ronny Thomale, Steven Kivelson, Harshit Agarwal, Stuart Brown, Hans-Joachim Elmers, Olena Fedchenko, Mark Fischer, Zurab Guguchia, Amir A. Haghighirad, Matthieu Le Tacon, Titus Neupert, Eduardo H. da Silva Neto, and Roser Valentí for stimulating discussions. 
\end{acknowledgements}

\textbf{Funding:} D.J.S. and J.S. disclose support for the research of this work from the German Research Foundation (DFG) through CRC TRR 288 ``Elasto-Q-Mat,'' project A07. J.S. discloses support for the research of this work from the Simons Foundation Collaboration on New Frontiers in Superconductivity (Grant SFI-MPS-NFS-00006741-03). Y.B.K. discloses support for the research of this work from the Natural Science and Engineering Research Council of Canada (NSERC) Discovery Grant No. RGPIN-2023-03296 and the Center for Quantum Materials at the University of Toronto. R.M.F. discloses support for the research of this work from the Air Force Office of Scientific Research under Award No. FA9550-21-1-0423, and a Mercator Fellowship from the German Research Foundation
(DFG) through CRC TRR 288, 422213477 ``Elasto-Q-Mat.''

\clearpage

\onecolumngrid

\begin{center}
\textbf{\large Supplementary to ``Superconductivity in kagome metals due to soft loop-current fluctuations"}
\end{center}
\begin{center}
Daniel J. Schultz$^1$, Grgur Palle$^{2,3}$, Asimpunya Mitra$^4$, Yong Baek Kim$^4$, Rafael M. Fernandes$^{2,3}$, and J\"org Schmalian$^{1,5}$
\end{center}
\begin{center}
{\small
\textit{$^1$Institute for Theoretical Condensed Matter Physics,
Karlsruhe Institute of Technology, 76131 Karlsruhe, Germany}

\textit{$^2$Department of Physics, The Grainger College of Engineering, University of Illinois Urbana-Champaign, Urbana, Illinois 61801, USA}

\textit{$^3$Anthony J. Leggett Institute for Condensed Matter Theory, The Grainger College of Engineering, University of Illinois Urbana-Champaign, Urbana, Illinois 61801, USA}

\textit{$^4$Department of Physics, University of Toronto, Toronto, Ontario M5S 1A7, Canada}

\textit{$^5$Institute for Quantum Materials and Technologies, Karlsruhe Institute of Technology, 76131 Karlsruhe, Germany}}
\end{center}


\section{(Reduced) space group transformations} \label{app:space_group_transformations}
\begin{figure}[h]
\centering
\includegraphics[scale=1]{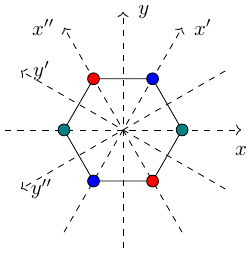}
\caption{In-plane rotation axes for the $C_2$ rotations. The hexagon orientation pictured here is the same hexagon appearing in the kagome lattice.} \label{fig:rotation_axes_hexagon}
\end{figure}

\ce{AV3Sb5} has the space group P6/mmm. This is a symmorphic group, and thus may be written as a semidirect product of the $D_{6h}$ point group with the group of lattice translations. Since we are focusing on a single layer, we ignore translations in the $z$ (out of plane) direction. We define the $x$ and $y$ (and other equivalent) rotation axes in Fig.~\ref{fig:rotation_axes_hexagon}. The $z$-axis points out of the page.

\begin{figure}[h]
\centering
\includegraphics[scale=1]{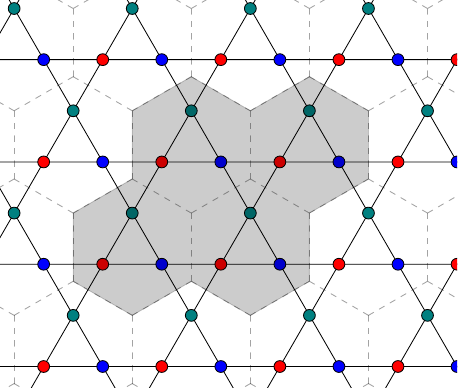}
\caption{Extended unit cell. All loop current patterns live within the greyed out area, and are repeated periodically around the lattice using the translation vectors $2\mathbf{a}_1$ and $2\mathbf{a}_2$. Note that the Sb atoms are not depicted in this figure. Only the V atom lattice sites are shown here.} \label{fig:extended_unitcell}
\end{figure}

The group theory classification in this work concerns loop current patterns living in the $2\times 2$ extended unit cell, as pictured in Fig.~\ref{fig:extended_unitcell}. We therefore consider the irreducible representations of a \textit{reduced space group} (sometimes also called extended point group) consisting of four unit cells. These four unit cells contain 12 kagome lattice sites (because there are 3 kagome/vanadium sites per unit cell), and 24 nearest neighbour bonds. In addition to the 24 point group operations of $D_{6h}$, we may now perform translations by $\bm{a}_1 = (1,0)$, $\bm{a}_2 = (\frac{1}{2},\frac{\sqrt{3}}{2})$, and $\bm{a}_1 + \bm{a}_2$. Within this reduced space group, a translation by $2\bm{a}_1$ or $2\bm{a}_2$ is the same as the identity operation. The result is a group of 96 elements. The character table and conjugacy classes for this group are outlined in Table \ref{tab:reduced_space_group}. We remark that $\sigma_h$ is the operation which sends $z \to -z$, and the other mirror reflections are $\sigma_v: y \to -y$ and $\sigma_d: x \to -x$. These other mirror reflections may alternatively be written as $\sigma_v = I\circ C_{2y}$ and $\sigma_d = I\circ C_{2x}$.

\begin{table}
\centering
\begin{tabular}{c|cccccc|cccccc|cccc|cccc|c}
Irrep & $E$ & $C_{6z}$ & $C_{3z}$ & $C_{2z}$         & $C_{2x}$          & $C_{2y}$         & $I$ & $S_6$ & $S_3$ & $\sigma_h$          & $\sigma_v$         & $\sigma_d$ & $T$ & $C^t_{2z}$         & $C^t_{2x}$          & $C^t_{2y}$         & $I^t$ & $\sigma^t_h$          & $\sigma^t_v$         & $\sigma^t_d$ & Basis functions \\ \hline
$A_{1g}$ &  1 & 1 & 1 & 1 & 1 & 1 & 1 & 1 & 1 & 1 & 1 & 1 & 1 & 1 & 1 & 1 & 1 & 1 & 1 & 1 & $\{\ket{z^2}\}$ \\
$A_{2g}$ & 1 & 1 & 1 & 1 & $-1$ & $-1$ & 1 & 1 & 1 & 1 & $-1$ & $-1$ & 1 & 1 & $-1$ & $-1$ & 1 & 1 & $-1$ & $-1$ & $\{\ket{xy(x^2-3y^2)(3x^2-y^2)}\}$\\
$B_{1g}$ & 1 & $-1$ & 1 & $-1$ & 1 & $-1$ & 1 & 1 & $-1$ & $-1$ & $-1$ & 1 & 1 & $-1$ & 1 & $-1$ & 1 & $-1$ & $-1$ & 1 & $\{\ket{yz(x^2-3y^2)}\}$ \\
$B_{2g}$ & 1 & $-1$ & 1 & $-1$ & $-1$ & 1 & 1 & 1 & $-1$ & $-1$ & 1 & $-1$ & 1 & $-1$ & $-1$ & 1 & 1 & $-1$ & 1 & $-1$ & $\{\ket{zx(3x^2-y^2)}\}$ \\
$E_{1g}$ & 2 & 1 & $-1$ & $-2$ & 0 & 0 & 2 & $-1$ & 1 & $-2$ & 0 & 0 & 2 & $-2$ & 0 & 0 & 2 & $-2$ & 0 & 0 & $\{\ket{zx},\ket{yz}\}$ \\
$E_{2g}$ & 2 & $-1$ & $-1$ & 2 & 0 & 0 & 2 & $-1$ & $-1$ & 2 & 0 & 0 & 2 & 2 & 0 & 0 & 2 & 2 & 0 & 0 & $\{\ket{x^2-y^2},\ket{xy}\}$ \\
$A_{1u}$ & 1 & 1 & 1 & 1 & 1 & 1 & $-1$ & $-1$ & $-1$ & $-1$ & $-1$ & $-1$ & 1 & 1 & 1 & 1 & $-1$ & $-1$ & $-1$ & $-1$  & $\{\ket{xyz(x^2-3y^2)(3x^2-y^2)}\}$\\
$A_{2u}$ & 1 & 1 & 1 & 1 & $-1$ & $-1$ & $-1$ & $-1$ & $-1$ & $-1$ & 1 & 1 & 1 & 1 & $-1$ & $-1$ & $-1$ & $-1$ & 1 & 1 & $\{\ket{z}\}$ \\
$B_{1u}$ & 1 & $-1$ & 1 & $-1$ & 1 & $-1$ & $-1$ & $-1$ & 1 & 1 & 1 & $-1$ & 1 & $-1$ & 1 & $-1$ & $-1$ & 1 & 1 & $-1$ & $\{\ket{x(x^2-3y^2)}\}$ \\
$B_{2u}$ & 1 & $-1$ & 1 & $-1$ & $-1$ & 1 & $-1$ & $-1$ & 1 & 1 & $-1$ & 1 & 1 & $-1$ & $-1$ & 1 & $-1$ & 1 & $-1$ & 1 & $\{\ket{y(3x^2-y^2)}\}$ \\
$E_{1u}$ & 2 & 1 & $-1$ & $-2$ & 0 & 0 & $-2$ & 1 & $-1$ & 2 & 0 & 0 & 2 & $-2$ & 0 & 0 & $-2$ & 2 & 0 & 0 & $\{\ket{x},\ket{y}\}$ \\
$E_{2u}$ & 2 & $-1$ & $-1$ & 2 & 0 & 0 & $-2$ & 1 & 1 & $-2$ & 0 & 0 & 2 & 2 & 0 & 0 & $-2$ & $-2$ & 0 & 0 & $\{\ket{z(x^2-y^2)},\ket{xyz}\}$ \\ \hline
$M_1^+$ & 3 & 0 & 0 & 3 & 1 & 1 & 3 & 0 & 0 & 3 & 1 & 1 & $-1$ & $-1$ & $-1$ & $-1$ & $-1$ & $-1$ & $-1$ & $-1$ \\
$M_2^+$ & 3 & 0 & 0 & 3 & $-1$ & $-1$ & 3 & 0 & 0 & 3 & $-1$ & $-1$ & $-1$ & $-1$ & 1 & 1 & $-1$ & $-1$ & 1 & 1 \\
$M_3^+$ & 3 & 0 & 0 & $-3$ & 1 & $-1$ & 3 & 0 & 0 & $-3$ & $-1$ & 1 & $-1$ & 1 & $-1$ & 1 & $-1$ & 1 & 1 & $-1$ \\
$M_4^+$ & 3 & 0 & 0 & $-3$ & $-1$ & 1 & 3 & 0 & 0 & $-3$ & 1 & $-1$ & $-1$ & 1 & 1 & $-1$ & $-1$ & 1 & $-1$ & 1 \\
$M_1^-$ & 3 & 0 & 0 & 3 & 1 & 1 & $-3$ & 0 & 0 & $-3$ & $-1$ & $-1$ & $-1$ & $-1$ & $-1$ & $-1$ & 1 & 1 & 1 & 1 \\
$M_2^-$ & 3 & 0 & 0 & 3 & $-1$ & $-1$ & $-3$ & 0 & 0 & $-3$ & 1 & 1 & $-1$ & $-1$ & 1 & 1 & 1 & 1 & $-1$ & $-1$ \\
$M_3^-$ & 3 & 0 & 0 & $-3$ & 1 & $-1$ & $-3$ & 0 & 0 & 3 & 1 & $-1$ & $-1$ & 1 & $-1$ & 1 & 1 & $-1$ & $-1$ & 1 \\
$M_4^-$ & 3 & 0 & 0 & $-3$ & $-1$ & 1 & $-3$ & 0 & 0 & 3 & $-1$ & 1 & $-1$ & 1 & 1 & $-1$ & 1 & $-1$ & 1 & $-1$ \\
\end{tabular}
\caption{Character table of the reduced space group, corresponding to $D_{6h}$ plus translations modulo $2\bm{a}_1$ and $2\bm{a}_2$. Note that, if we are talking about fermion bilinears (like loop current operators), then we can also classify the operators according to their character under time-reversal, and we insert an $m$, e.g. $mA_{2g}$ to denote a time-reversal odd irrep, which spatially transforms like $A_{2g}$.} \label{tab:reduced_space_group}
\end{table}

\clearpage

\section{Current operators and the current representation}
\subsection{Derivation of current operator}
The local current operator for a charge-conserving (global $\text{U}(1)$ phase-rotation symmetric), time-reversal symmetric, lattice model can be found by considering the number density operator $\hat{n}_{i} = \sum_{a\sigma}\hat{c}^\dagger_{ia\sigma}\hat{c}_{ia\sigma}$, whereby $i$ is a lattice site, and $a$ labels orbitals on that site. The time-dependent operator $\hat{n}_{i}(t) = e^{i\hat{H} t}\hat{n}_{i} e^{-i\hat{H} t}$ satisfies the Heisenberg equation of motion 
\begin{align}
\frac{\text{d} \hat{n}_{i}(t)}{\text{d} t} ={}& i[\hat{H},\hat{n}_{i}(t)]. 
\end{align}
We then use the conservation law $\text{d}\hat{n}_{i}/\text{d} t + (\nabla\cdot\bm{j})_i = 0$. Note that the discretized divergence is given by 
\begin{equation}
(\nabla\cdot\hat{\bm{j}})_i = -\sum_j \hat{j}_{ij}, \label{eq:current_operator}
\end{equation}
where $\hat{j}_{ij}$ represents the current flowing from site $j$ to site $i$. Consider a free fermion system described by a hopping Hamiltonian of the form
\begin{align}
\hat{H} = \sum_{ijab\sigma} \hat{c}^\dagger_{ia\sigma} H_{ijab} \hat{c}_{jb\sigma},
\end{align}
which must satisfy the relation $H_{ijab} = H_{jiba}^*$ due to Hermiticity. Here, $a,b$ label the different orbitals on the sites $i,j$ respectively. By computing the commutator $[\hat{H},\hat{n}_{i\sigma}]$, the form of the current operator follows
\begin{equation}
\hat{j}_{ij} = i\sum_{ab\sigma} (-H_{ijab} \hat{c}^\dagger_{ia\sigma} \hat{c}_{jb\sigma} + H_{jiba} \hat{c}^\dagger_{jb\sigma} \hat{c}_{ia\sigma}).
\end{equation}
We note that the current operator is odd under swapping the two sites and that it is Hermitian, $\hat{j}_{ij} = -\hat{j}_{ji} = \hat{j}_{ij}^\dagger$. Furthermore, it is odd under the time-reversal operation: $\hat{\mathcal{T}}\hat{j}_{ij}\hat{\mathcal{T}}^{-1} = -\hat{j}_{ij}$. Group theoretical considerations alone are not enough to construct the valid current patterns. Some of the current patterns in the group theory classification are not allowed, due to the following two constraints:
\begin{enumerate}
\item Kirchoff constraint: there can be no net current into/out of a particular site. Such a scenario would result in charge buildup on that site. In mathematical terms, $\langle (\nabla\cdot \hat{\bm{j}})_i\rangle = 0$, for every site $i$.
\item Bloch constraint: there can be no global current. In mathematical terms, if we define the net current (which is a vector) through a site by $\hat{\bm{J}}_i = \sum_j \bm{r}_{ij} \hat{j}_{ij}$ (whereby $\bm{r}_{ij} = \frac{\bm{R}_i - \bm{R}_j}{|\bm{R}_i - \bm{R}_j|}$), then the total current must vanish $\sum_i \langle \hat{\bm{J}}_i\rangle = 0$. This condition is automatically satisfied for any loop current pattern that has a nonzero ordering wave vector.
\end{enumerate}
Some of the loop current patterns generated from the group theoretical analysis violate these constraints, and must therefore be discarded. All patterns that we have presented in the main text obey these two constraints. 

\subsection{The current representation}
In our model, we have a variety of hoppings into/out of a kagome lattice site, as well as into/out of the in-plane Sb site. There are nearest and next-nearest V hoppings, as well as hybridizations between V and Sb, both in and out of plane. This leads to a huge number of bonds along which hopping can occur, and allowing currents to flow among all of them greatly complicates our analysis. For simplicity, we consider two classes of current patterns:
\begin{enumerate}
\item Current patterns that flow only between nearest neighbor V sites. There are 24 such bonds, hence creating a 24 dimensional vector space to represent the reduced space group on. As mentioned in the main text, the tight-binding model can be separated into the $\sigma_h=1$ and $\sigma_h=-1$ sectors. Since we are considering only $\sigma_h = -1$, we allow only currents between $d_{zx},d_{yz}$ orbitals. This representation is then reduced as follows:
\begin{align}
R_{\text{V-V current}} = mA_{2g} \oplus mE_{2g} \oplus mB_{1u} \oplus mE_{1u} \oplus mM_{1}^+ \oplus mM_{2}^+ \oplus mM_{2}^+ \oplus mM_{3}^- \oplus mM_{3}^- \oplus mM_{4}^- 
\end{align}
The $m$ prefix indicates oddness under time-reversal.
For the first four irreps that have zero momentum (Table \ref{tab:reduced_space_group}), the $g/u$ subscripts indicate evenness/oddness under spatial inversion, respectively, while for the last six $M$-point irreps, the $\pm$ superscripts denote the evenness/oddness under spatial inversion. Both copies of $mM_{2}^+$ are analyzed in the main text, namely patterns (a) and (b) of Fig.~3. Although there are two copies of $mM_{3}^-$, only one fine-tuned linear combination of them satisfies the Kirchhoff constraint and is pictured in pattern (d) of Fig.~3.
\item Current patterns that flow between nearest neighbor V and in-plane Sb sites. There are also 24 such bonds, hence creating a 24 dimensional vector space to represent the reduced space group on. Since we are considering only $\sigma_h = -1$, we only allow currents between $d_{zx},d_{yz}$ and $p_z$ (from planar Sb) orbitals.
The decomposition:
\begin{align}
R_{\text{V-Sb current}} = mA_{1g} \oplus mE_{2g} \oplus mB_{1u} \oplus mE_{1u} \oplus mM_{1}^+ \oplus mM_{1}^+ \oplus mM_{2}^+ \oplus mM_{3}^- \oplus mM_{3}^- \oplus mM_{4}^- 
\end{align}
The one copy of $mM_{2}^+$ is shown in the main text, Fig.~3, as pattern (c).
The pattern (e) of Fig.~3 is made of $mM_{3}^-$ that is superimposed (in a way satisfying the Kirchhoff constraint) with a $mM_{3}^-$ pattern coming from V-V patterns.
\end{enumerate}

\subsection{Current patterns}
\begin{figure}
\centering
\includegraphics[scale=1]{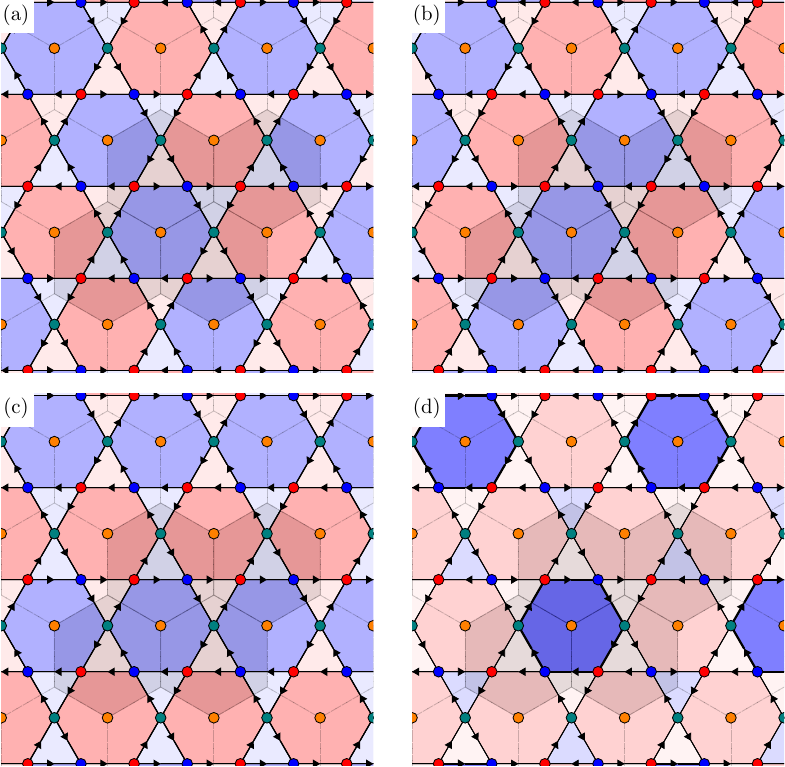}
\caption{The current patterns for Fig.~3(a) of the main text. They are in the irrep $mM^+_2$ and only have currents flowing between V-V. The patterns (a),(b),(c) have ordering wave vectors $\bm{M}_1$, $\bm{M}_2$, and $\bm{M}_3$, respectively. The fourth pattern (d) is the superposition of the three patterns.} \label{fig:LC_1}.
\end{figure}

\begin{figure}
\centering
\includegraphics[scale=1]{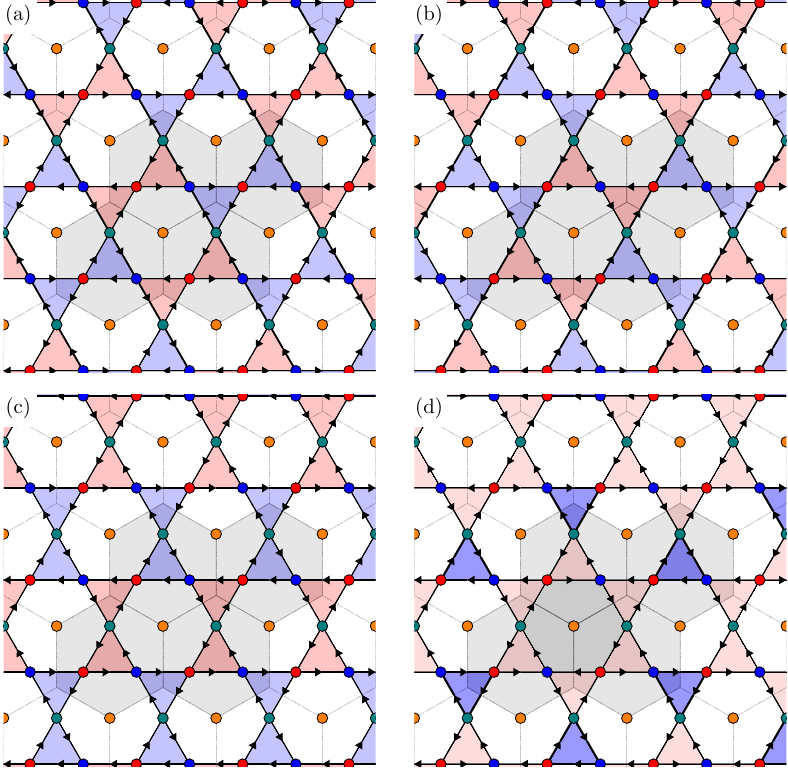}
\caption{The current patterns for Fig.~3(b) of the main text. They are in the irrep $mM^+_2$ and only have currents flowing between V-V. The patterns (a),(b),(c) have ordering wave vectors $\bm{M}_1$, $\bm{M}_2$, and $\bm{M}_3$, respectively. The fourth pattern (d) is the superposition of the three patterns.} \label{fig:LC_2}
\end{figure}

\begin{figure}
\centering
\includegraphics[scale=1]{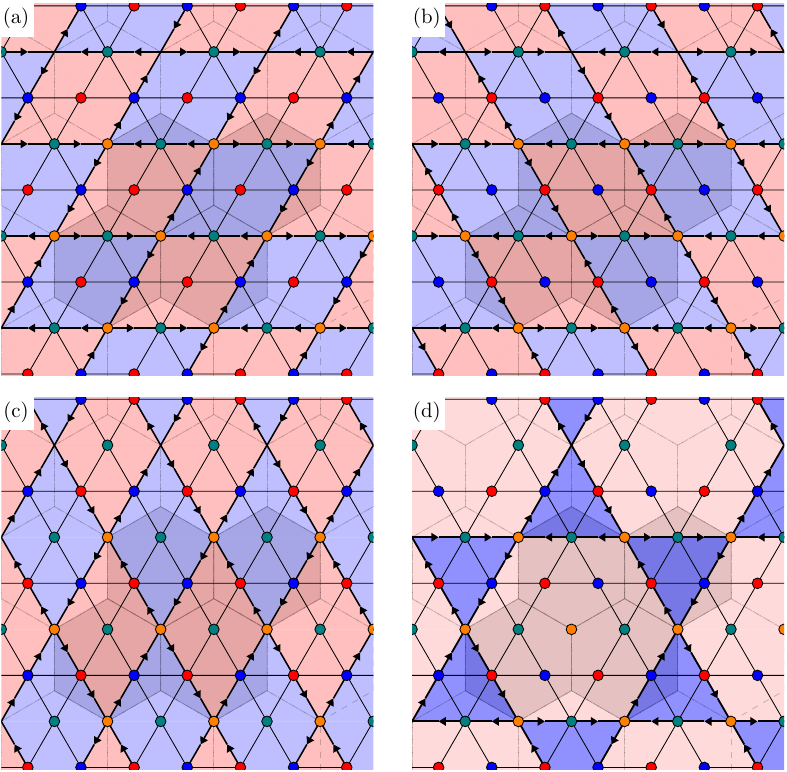}
\caption{The current patterns for Fig.~3(c) of the main text. They are in the irrep $mM^+_2$ and only have currents flowing between V-Sb. The patterns (a),(b),(c) have ordering wave vectors $\bm{M}_1$, $\bm{M}_2$, and $\bm{M}_3$, respectively. The fourth pattern (d) is the superposition of the three patterns.} \label{fig:LC_3}
\end{figure}

\begin{figure}
\centering
\includegraphics[scale=1]{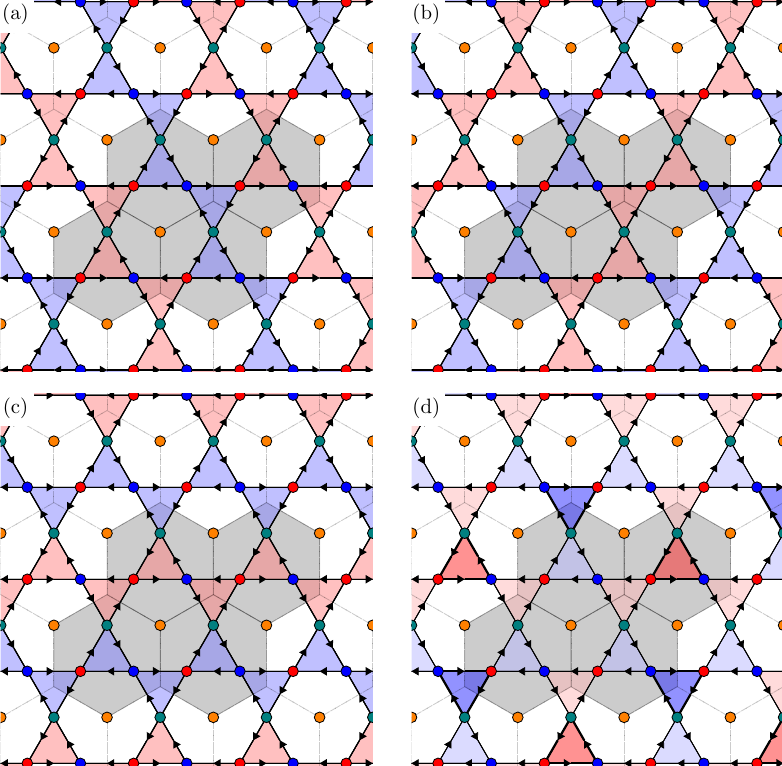}
\caption{The current patterns for Fig.~3(d) of the main text. They are in the irrep $mM^-_3$ and only have currents flowing between V-V. The patterns (a),(b),(c) have ordering wave vectors $\bm{M}_1$, $\bm{M}_2$, and $\bm{M}_3$, respectively. The fourth pattern (d) is the superposition of the three patterns.} \label{fig:LC_4}
\end{figure}

\begin{figure}
\centering
\includegraphics[scale=1]{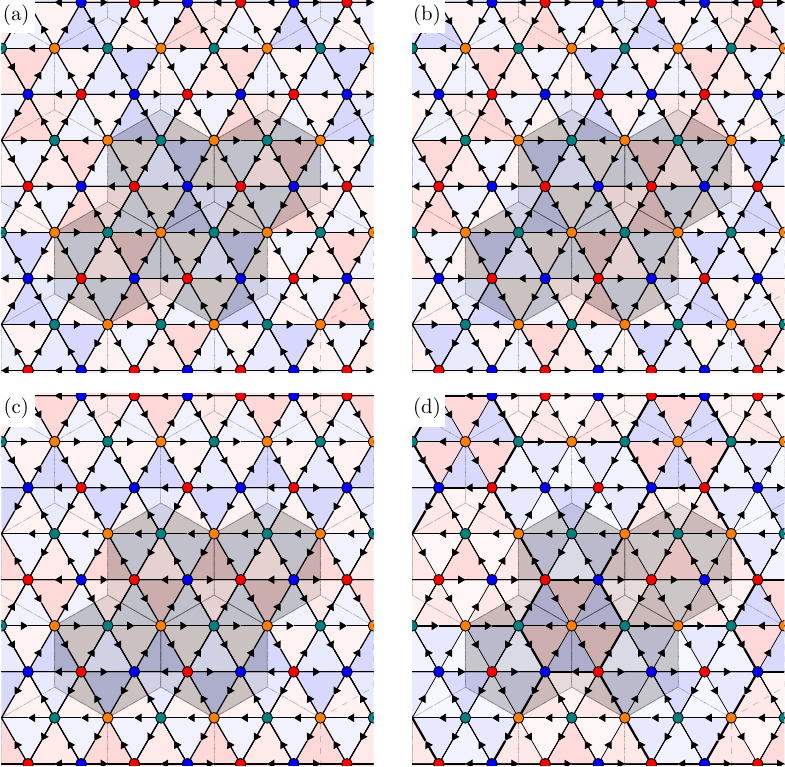}
\caption{The current patterns for Fig.~3(e) of the main text. They are in the irrep $mM^-_3$ and have both currents flowing between V-V and V-Sb. The patterns (a),(b),(c) have ordering wave vectors $\bm{M}_1$, $\bm{M}_2$, and $\bm{M}_3$, respectively. The fourth pattern (d) is the superposition of the three patterns.} \label{fig:LC_5}
\end{figure}

In this section, we discuss in more detail the current patterns listed in the main text and what is their relationship to other current patterns discussed in the literature. Because all current patterns have one of the $M$-points as their ordering wave vector, they come in sets of three: one with ordering wave vector $\bm{M}_1$, one with ordering wave vector $\bm{M}_2$, and one with ordering wave vector $\bm{M}_3$; see panel (f) of Fig.~3 of the main text.
Only the $\bm{M}_3$ representative is shown in the main text, but here in Figs.~\ref{fig:LC_1} to~\ref{fig:LC_5} we also show the other representatives, as well as the corresponding 3-$\bm{Q}$ orders that are equal-weight $\propto (1, 1, 1)$ superposition of the three individual 1-$\bm{Q}$ orders.

For the reader's convenience, below we identify the patterns considered by us with some selected works from the literature.
\begin{itemize}
\item The pattern in Fig.~2c of Ref.~\cite{feng_chiral_2021} is in the irrep $mM^+_2$, and is a weighted superposition of our patterns in Figs.~\ref{fig:LC_1} and Figs.~\ref{fig:LC_2}.
\item The patterns in Fig.~2 of Ref.~\cite{yang_intertwining_2023} are all in the irrep $mM^+_2$, and are a weighted superposition of our patterns in Figs.~\ref{fig:LC_1} and Figs.~\ref{fig:LC_2}.
\item The four patterns in Fig.~4b--4e of Ref.~\cite{dong_loop-current_2023} are all in the irrep $mM^+_2$. In particular, they are all weighted superpositions of our patterns in Figs.~\ref{fig:LC_1} and \ref{fig:LC_2}, and their Fig.~4d even appears to be the specific linear combination comprising only our Fig.~\ref{fig:LC_2}.
\item There are four patterns in Fig.~6 of Ref.~\cite{feng_low-energy_2021}. In their Fig.~6a (irrep $mM^+_2$), this is a weighted superposition of our patterns in Figs.~\ref{fig:LC_1} and Figs.~\ref{fig:LC_2}. In their Fig.~6b (irrep $mM^+_2$), this is our Fig.~\ref{fig:LC_1}. In their Fig.~6c (irrep $mM^+_2$), this is a weighted superposition of our patterns in Figs.~\ref{fig:LC_1} and Figs.~\ref{fig:LC_2}. In their Fig.~6d (irrep $mM^-_3$), this is our Fig.~\ref{fig:LC_4}.
\item The pattern in Fig.~2a of Ref.~\cite{li_intertwined_2024} is in the irrep $mM^+_2$, and is a weighted superposition of our patterns in Figs.~\ref{fig:LC_1} and Figs.~\ref{fig:LC_2}.
\end{itemize}

\clearpage
\section{Phenomenological vs.\ microscopic susceptibilities}
In this section, we microscopically calculated loop current propagators for various loop current patterns using the random phase approximation (RPA) and compare them with the phenomenological form written down in Eq.~(3) of the main text. Both give very similar results for the pairing states, as determined by the linearized gap equation.

\subsection{Microscopic computation of loop-current susceptibility}
For the computation of a loop current susceptibility, we consider a loop current pattern defined by a fermion bilinear 
\begin{align}
\hat{J}^\ell_{\bm{q}} ={}& \sum_{\bm{k} ab\sigma} \hat{c}^\dagger_{\bm{k}, a,\sigma} J^\ell_{ab}(\bm{k},\bm{k}+\bm{q}) \hat{c}_{\bm{k}+\bm{q},b,\sigma}
\end{align}
Here, $\ell \in \{1, 2, 3\}$ labels the three different current patterns within the $M$-point irrep. The definition of the loop current susceptibility is ($n$ labels the unit cell):
\begin{align}
\chi^{\ell\ell'}_{\text{LC}}(\bm{R}_n,\tau;\bm{R}_{n'}) ={}& -\langle T_\tau[\hat{J}^\ell_{n}(\tau) \hat{J}_{n'}]\rangle,
\end{align}
which, in Fourier space, reads
\begin{align}
\chi^{\ell\ell'}_{\text{LC}}(\bm{q},iq_0) ={}& -\frac{1}{N}\int_0^\beta e^{iq_0\tau} \langle T_\tau[\hat{J}^\ell_{\bm{q}}(\tau)\hat{J}_{-\bm{q}}]\rangle \d\tau \\
={}& -\frac{1}{\beta N}\langle J^\ell_{q} J^{\ell'}_{-q} \rangle
\end{align}
whereby the second line is written in the path integral language, such that the imaginary time ordering is automatic. If computed with respect to the noninteracting Hamiltonian, this gives the bare susceptibility. To go beyond the bare susceptibility, we need to include corrections generated by the interactions between loop current modes. The minimal approach to do this is through RPA. For this purpose, we add to our Hamiltonian the following bare local loop current interaction:
\begin{align}
\hat{H}_{\text{int}} = -\frac{1}{2NV}\sum_{\bm{q},\ell} :\hat{J}^\ell_{\bm{q}} \hat{J}^\ell_{-\bm{q}}:
\end{align}
One way of arriving at this interaction is by decomposing extended Hubbard interactions into various channels and then retaining only the relevant loop current channel that, by assumption, becomes dominant in our system.
Since our current operators $\hat{J}^\ell_{\bm{q}}$ have units of energy, proportional to the hopping elements, we put the coupling constant $V$ in the denominator. Hence, the effective interaction in this channel is $H_{\text{int}} \propto \text{hopping}^2/V$. Moving to the path integral formalism, we can write an imaginary time action for the theory:
\begin{align}
S ={}& \underbrace{\int_0^\beta \sum_{\bm{k} ab\sigma} \bar{c}_{\bm{k}\tau a\sigma} (\delta_{ab}\partial_\tau + H_{ab}(\bm{k})) c_{\bm{k}\tau b\sigma} \d\tau}_{S_0} +\underbrace{ \frac{-1}{2NV}\int_0^\beta \sum_{\bm{q},\ell} J^\ell_{\bm{q}} J^\ell_{-\bm{q}} \d\tau }_{S_{\text{int}}} \\
={}& \underbrace{\sum_{k ab\sigma} \bar{c}_{k a\sigma} (-ik_0\delta_{ab} + H_{ab}(\bm{k})) c_{k b\sigma}}_{S_0} +\underbrace{\frac{-1}{2\beta NV}\sum_{q,\ell} J^\ell_{q} J^\ell_{-q}}_{S_{\text{int}}}
\end{align}

We now recall the Hubbard-Stratonovich transformation:
\begin{align}
\int \exp\left(-\frac{1}{\beta NV}\sum_{q}\left[\frac{1}{2}  |\Phi_{q}|^2 - \Phi_{q} J_{-q}\right]\right) \mathcal{D}\Phi = \exp\left(\frac{1}{2\beta NV}\sum_{q} J_{q} J_{-q}\right).
\end{align}
Here, $\Phi_q = \Phi_{-q}^*$ is the real bosonic field of the loop current mode. This allows us to rewrite the grand partition function as follows:
\begin{align}
\mathcal{Z} ={}& \int e^{-S_0 - S_{\text{int}}} \D[\bar{c},c] \\
={}& \int e^{-S_0} \exp\left(\frac{1}{2\beta NV} \sum_{q\ell} J^{\ell}_{q} J^{\ell}_{-q} \right) \D[\bar{c},c] \\
={}& \int e^{-S_0} \exp\left(-\frac{1}{\beta NV}\sum_{q\ell}\left[\frac{1}{2}|\Phi^\ell_q|^2 - \Phi^\ell_q J^\ell_{-q}\right]\right) \D\Phi\D[\bar{c},c]
\end{align}

The loop current susceptibility in the $J$-language is a 4-fermion correlation function that is directly related to a 2-point correlation in the $\Phi$-language. To show this, we write the correlation function as a functional derivative:
\begin{align}
\chi^{\ell\ell'}_{\text{LC}}(q) ={}& -\frac{1}{\beta N}\langle J^\ell_{q} J^{\ell'}_{-q} \rangle \\
={}& -\frac{1}{\beta N\mathcal{Z}}\int e^{-S_0} \exp\left(-\frac{1}{2VN\beta} \sum_{q\ell} |\Phi^\ell_{q}|^2\right) \exp\left(\frac{1}{\beta NV} \sum_{q\ell} \Phi^\ell_{q} J^\ell_{-q}\right)  J^\ell_{q} J^{\ell'}_{-q} \D\Phi\D[\bar{c},c] \\
={}& -\frac{1}{\mathcal{Z}}\int e^{-S_0} \exp\left(-\frac{1}{2VN\beta} \sum_{q\ell} |\Phi^\ell_{q}|^2\right) \left[\beta NV^2\frac{\delta}{\delta \Phi^\ell_{-q}} \frac{\delta}{\delta \Phi^{\ell'}_{q}} \exp\left(\frac{1}{\beta NV} \sum_{q\ell} \Phi^\ell_{q} J^\ell_{-q}\right) \right] \D\Phi\D[\bar{c},c]
\end{align}
Through two integrations by parts, this can be rewritten as
\begin{align}
\chi^{\ell\ell'}_{\text{LC}}(q) ={}& -\frac{1}{\mathcal{Z}}\int e^{-S_0} \left(-V\delta_{\ell\ell'} + \frac{1}{\beta N} \Phi^{\ell'}_{-q}\Phi^{\ell}_{q}\right)\exp\left(-\frac{1}{2VN\beta} \sum_{q\ell} |\Phi^\ell_{q}|^2\right) \exp\left(\frac{1}{\beta NV} \sum_{q\ell} \Phi^\ell_{q} J^\ell_{-q}\right) \D\Phi\D[\bar{c},c] \\
={}& V\delta_{\ell\ell'} - \frac{1}{\beta N} \langle \Phi^\ell_{q}  \Phi^{\ell'}_{-q} \rangle \label{eq:LC-susceptibility-formula}
\end{align}

The RPA susceptibility is then obtained very easily by computing the $\langle \Phi^\ell_{q}  \Phi^{\ell'}_{-q} \rangle$ correlation function at leading order in the coupling. To do this, we integrate out the electrons and then truncate the effective action for $\Phi$ at quadratic order. Doing this, we find that
\begin{align}
\mathcal{Z} ={}& \int e^{-S_{\text{eff}}[\Phi]} \D\Phi \\
e^{-S_{\text{eff}}[\Phi]} ={}& \int e^{-S} \D[\bar{c},c] \\
={}& \exp\left(-\frac{1}{2VN\beta} \sum_{q\ell} |\Phi^\ell_{q}|^2\right)\int \exp\left(\sum_{kab} \bar{c}_{ka\sigma} [\mathcal{G}_0^{-1}(\bm{k},ik_0)]_{ab}c_{kb\sigma} + \frac{1}{\beta NV}\sum_{qk\ell ab\sigma} \Phi^\ell_{k-q} \bar{c}_{ka\sigma} J^\ell_{ab}(\bm{k},\bm{q}) c_{qb\sigma} \right)\D[\bar{c},c] \\
={}& \exp\left(-\frac{1}{2VN\beta} \sum_{q\ell} |\Phi^\ell_{q}|^2\right)\int \exp\left(\sum_{kqab\sigma} \bar{c}_{ka\sigma} \left(\delta_{kq}[\mathcal{G}_0^{-1}(\bm{k},ik_0)]_{ab} + \frac{1}{\beta NV}\sum_\ell \Phi^\ell_{k-q} J^\ell_{ab}(\bm{k},\bm{q})\right) c_{qb\sigma} \right)\D[\bar{c},c] \\
={}& \exp\left(-\frac{1}{2VN\beta} \sum_{q\ell} |\Phi^\ell_{q}|^2\right) \det\left(\delta_{kq}[\mathcal{G}_0^{-1}(\bm{k},ik_0)]_{ab} + \frac{1}{\beta NV}\sum_\ell \Phi^\ell_{k-q} J^\ell_{ab}(\bm{k},\bm{q})\right)
\end{align}
whereby we define the (inverse of the) noninteracting Green function to be 
\begin{align}
[\mathcal{G}_0^{-1}(\bm{k},ik_0)]_{ab} ={}& ik_0\delta_{ab} - H_{ab}(\bm{k}).
\end{align}
This allows us to expand the effective action in terms of powers of $\Phi$ and get that 
\begin{align}
S_{\text{eff}}[\Phi] ={}& \frac{1}{2V\beta N}\sum_{q\ell} |\Phi^\ell_q|^2 - \log\det \left(\delta_{kq}[\mathcal{G}_0^{-1}(\bm{k},ik_0)]_{ab} + \frac{1}{\beta NV}\sum_\ell \Phi^\ell_{k-q} J^\ell_{ab}(\bm{k},\bm{q})\right) \\
={}& \frac{1}{2V\beta N}\sum_{q\ell} |\Phi^\ell_q|^2 - \tr\log \left(\delta_{kq}[\mathcal{G}_0^{-1}(\bm{k},ik_0)]_{ab} + \frac{1}{\beta NV}\sum_\ell \Phi^\ell_{k-q} J^\ell_{ab}(\bm{k},\bm{q})\right) \\
={}& \frac{1}{2V\beta N}\sum_{q\ell} |\Phi^\ell_q|^2 - \tr\log \left(\delta_{kq}\delta_{ac} + \frac{1}{\beta NV}\sum_\ell \Phi^\ell_{k-q} [\mathcal{G}_0(\bm{k},ik_0)]_{ab} J^\ell_{bc}(\bm{k},\bm{q})\right) + \text{const},
\end{align}
whereby the constant arises after we factor out the noninteracting Green function. Using the standard Taylor expansion
\begin{align}
\log(\mathbb{1} + X) = X - \frac{X^2}{2} + \frac{X^3}{3} + \cdots,
\end{align}
we truncate the effective action at second order, and find that
\begin{align}
S^{(2)}_{\text{eff}}[\Phi] ={}& \frac{1}{2V\beta N}\sum_{q\ell} |\Phi^\ell_q|^2 + \frac{1}{2(\beta NV)^2}\sum_{\substack{kq\ell\ell' \\ aba'b'}} \Phi^\ell_{-q}\Phi^{\ell'}_{q}  [\mathcal{G}_0(\bm{k},ik_0)]_{ab} J^\ell_{ba'}(\bm{k},\bm{k}+\bm{q}) [\mathcal{G}_0(\bm{k}+\bm{q},ik_0+iq_0)]_{a'b'} J^{\ell'}_{b'a}(\bm{k}+\bm{q},\bm{k}) \\
={}& \frac{1}{2\beta NV} \sum_{q\ell\ell'} \Phi^\ell_{-q} [\mathcal{D}^{-1}_{\text{RPA}}(\bm{q},iq_0)]_{\ell\ell'} \Phi^{\ell'}_{q}.
\end{align}
This defines the (inverse of the) loop current boson's RPA-dressed propagator:
\begin{align}
[\mathcal{D}^{-1}_{\text{RPA}}(\bm{q},iq_0)]_{\ell\ell'} ={}& \underbrace{\delta_{\ell\ell'}}_{[\mathcal{D}^{-1}_0(\bm{q},iq_0)]_{\ell\ell'}} + \frac{1}{\beta NV}\sum_{k} \tr [\mathcal{G}_0(\bm{k},ik_0) J^\ell(\bm{k},\bm{k}+\bm{q}) \mathcal{G}_0(\bm{k}+\bm{q},ik_0+iq_0) J^{\ell'}(\bm{k}+\bm{q},\bm{k})].
\end{align}
In terms of the eigenvectors $\ket{u_n(\bm{k})}$ and eigenvalues $\xi_n(\bm{k})$ of the Hamiltonian $H(\bm{k})$, the bare fermionic Green function is given by $\mathcal{G}_0(\bm{k},ik_0) = \sum_n (ik_0 - \xi_n(\bm{k}))^{-1} \ket{u_n(\bm{k})} \bra{u_n(\bm{k})}$.
After inserting this and evaluating the fermionic Matsubara sum, we find that
\begin{align}
[\mathcal{D}^{-1}_{\text{RPA}}(\bm{q},iq_0)]_{\ell\ell'} = \delta_{\ell\ell'} - \frac{1}{N V} \sum_{\bm{k} m n} M_{\ell\ell'}^{mn}(\bm{k},\bm{k}+\bm{q}) \frac{- (f_{\bm{k}, m} - f_{\bm{k}+\bm{q}, n})}{\xi_{m}(\bm{k}) - \xi_{n}(\bm{k}+\bm{q}) + i q_0},
\end{align}
where $M_{\ell\ell'}^{mn}(\bm{k},\bm{k}+\bm{q}) = \bra{u_m(\bm{k})}{J^{\ell}(\bm{k},\bm{k}+\bm{q})}\ket{u_n(\bm{k}+\bm{q})} \bra{u_n(\bm{k}+\bm{q})}{J^{\ell'}(\bm{k}+\bm{q},\bm{k})}\ket{u_m(\bm{k})}$ and $f_{\bm{k}, m} = [e^{\beta \xi_{m}(\bm{k})} + 1]^{-1}$ is the Fermi-Dirac distribution.
For $\ell = \ell'$, $M_{\ell\ell}^{mn}(\bm{k},\bm{k}+\bm{q}) \geq 0$ and the second term on the right-hand side is therefore always negative, pushing the dressed Green function towards divergence as we approach condensation of the corresponding loop current.
This RPA propagator for loop currents can now be inserted into the formula for the linearized gap equation, replacing the phenomenological propagator $g^2\mathcal{D}_{\text{LC}}(\bm{q}) \to \mathcal{D}_{\text{RPA}}(\bm{q})$ from all equations in the main text and also the ``Methods" section.

Lastly, let us note that the 2-point $\Phi$ correlation function at the RPA level equals:
\begin{align}
\langle \Phi^\ell_{q} \Phi^{\ell'}_{-q}\rangle ={}& \beta NV[\mathcal{D}_{\text{RPA}}(\bm{q},iq_0)]_{\ell\ell'}.
\end{align}
Plugging this into the formula~\eqref{eq:LC-susceptibility-formula} for the LC susceptibility, we obtain
\begin{align}
\chi^{\ell\ell'}_{\text{LC,RPA}}(q) ={}& V\delta_{\ell\ell'} - \frac{1}{\beta N} \beta NV[\mathcal{D}_{\text{RPA}}(\bm{q},iq_0)]_{\ell\ell'} \\
={}& V\delta_{\ell\ell'} - V[\mathcal{D}_{\text{RPA}}(\bm{q},iq_0)]_{\ell\ell'}.
\end{align}

\clearpage
\subsection{Plots of the phenomenological propagator vs.\ RPA propagator}
To compare the phenomenological LC propagator with the RPA propagator, we plot both in the entire Brillouin zone. For convenience, we repeat the phenomenological propagator from the main text, Eq.~(3), which was given by 
\begin{align}
[\mathcal{D}_{\text{LC}}(\bm{q})]_{\ell\ell'} ={}& \frac{\delta_{\ell\ell'}}{r + (1-r)f(\bm{q}-\bm{M}_\ell)}, \\
f(\bm{q}) ={}& \frac{2}{3} - \frac{2}{9}\big(\cos(\bm{q}\cdot\bm{a}_1) + \cos(\bm{q}\cdot\bm{a}_2) + \cos(\bm{q}\cdot\bm{a}_3)\big).
\end{align}
This propagator we directly compare with the RPA propagator from the previous section.
\begin{figure}[!ht]
\centering
\includegraphics[scale=0.9]{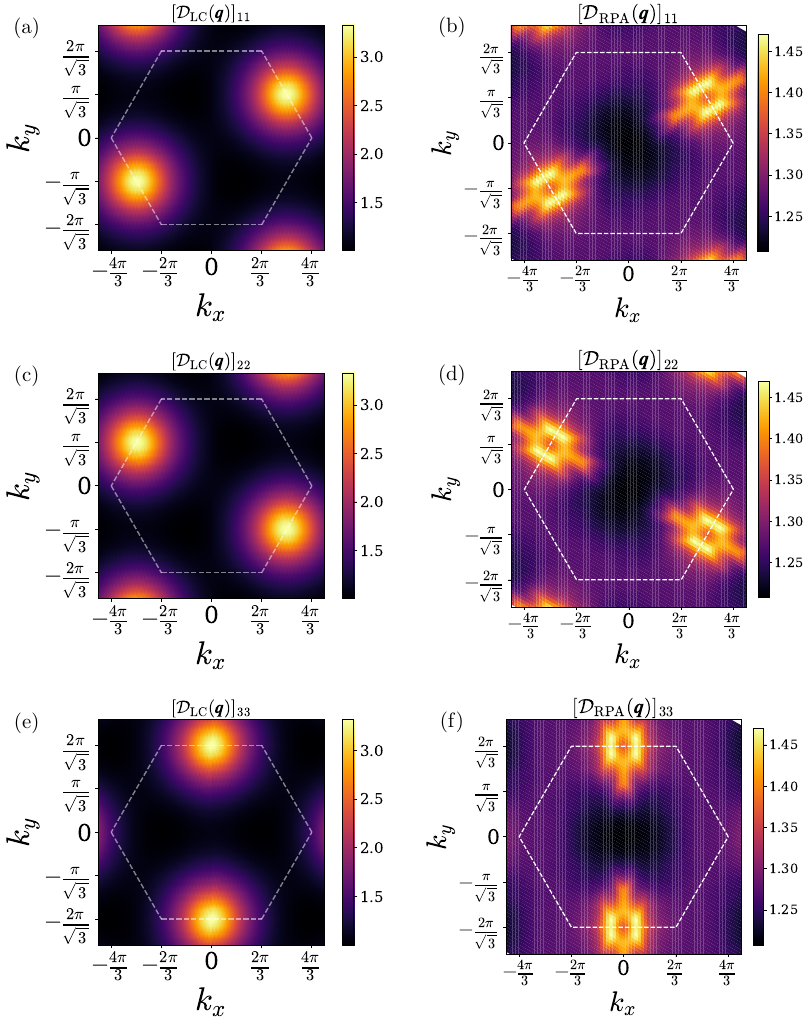}
\caption{Comparison of the phenomenological LC propagators (first column) with the RPA LC propagators (second column). In panels (a),(b), it is the susceptibility peaking at $\bm{M}_1$. Similarly for panels (c),(d) the peak is at $\bm{M}_2$ and for panels (e),(f) the peak is at $\bm{M}_3$. The peaks in the RPA propagators are not precisely at the $M$-points, but rather are closeby. This weak incommensurability is due to the imperfect nesting of the tight-binding model band structure. In spite of this difference, the pairing instabilities do not change.}
\end{figure}

\clearpage
\section{Projected Cooper interaction}
\begin{figure}[p]
\centering
\includegraphics[scale=0.95]{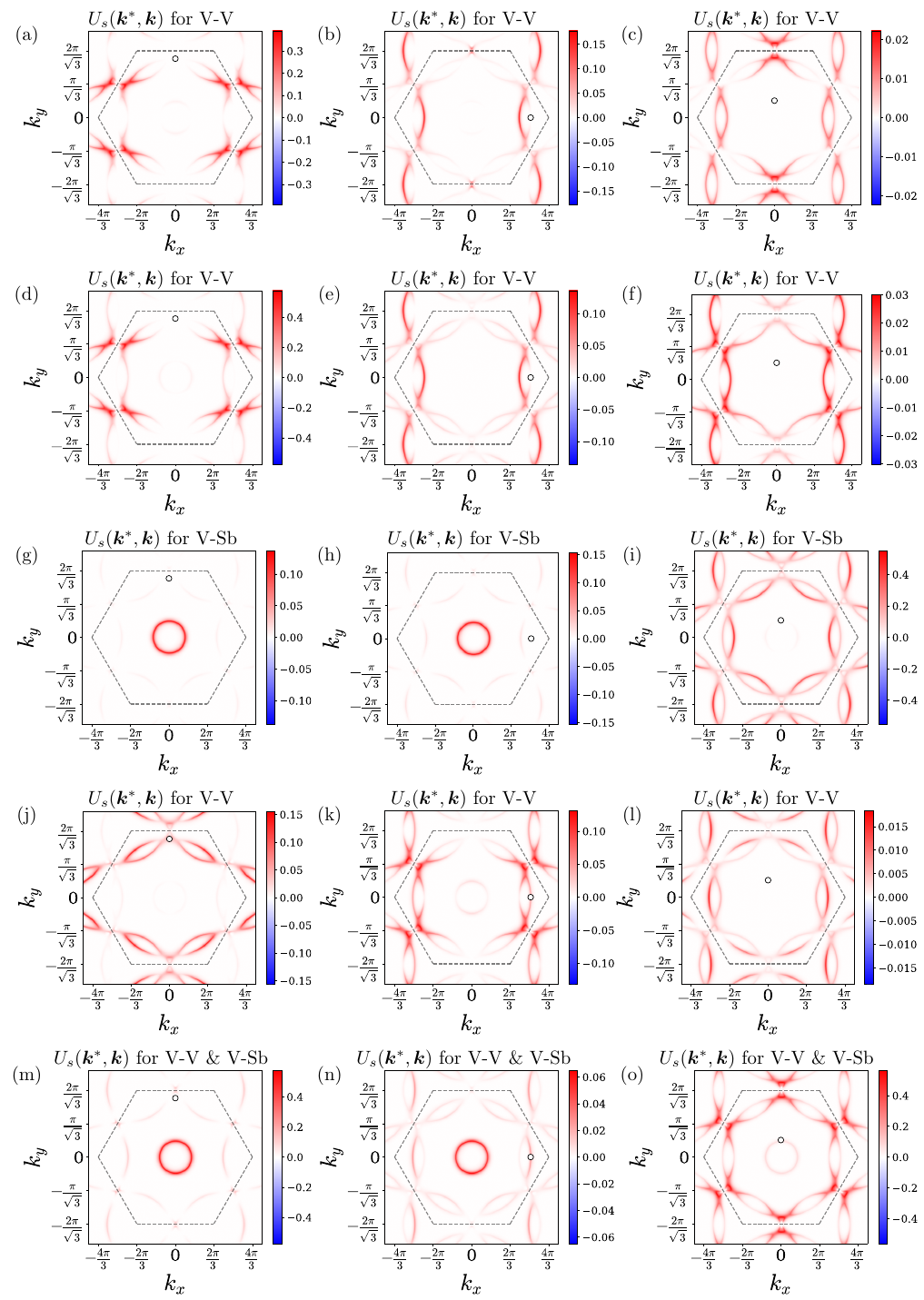}
\caption{Singet interactions. (a)-(c) are for main text Fig.~3(a). (d)-(f) are for main text Fig.~3(b). (g)-(i) are for main text Fig.~3(c). (j)-(l) are for main text Fig.~3(d). (m)-(o) are for main text Fig.~3(e). The key feature is that the first, second, and fourth rows (which are from V-V LC patterns) have very small interaction with the $\Gamma$ pocket, whereas lines 3 and 5 (which have a V-Sb component) have strong interaction with the $\Gamma$ pocket. This is the reason why the pathway is the determining factor for the resulting pairing symmetry.} \label{fig:all_singlet_interactions}
\end{figure}

\begin{figure}[p]
\centering
\includegraphics[scale=0.95]{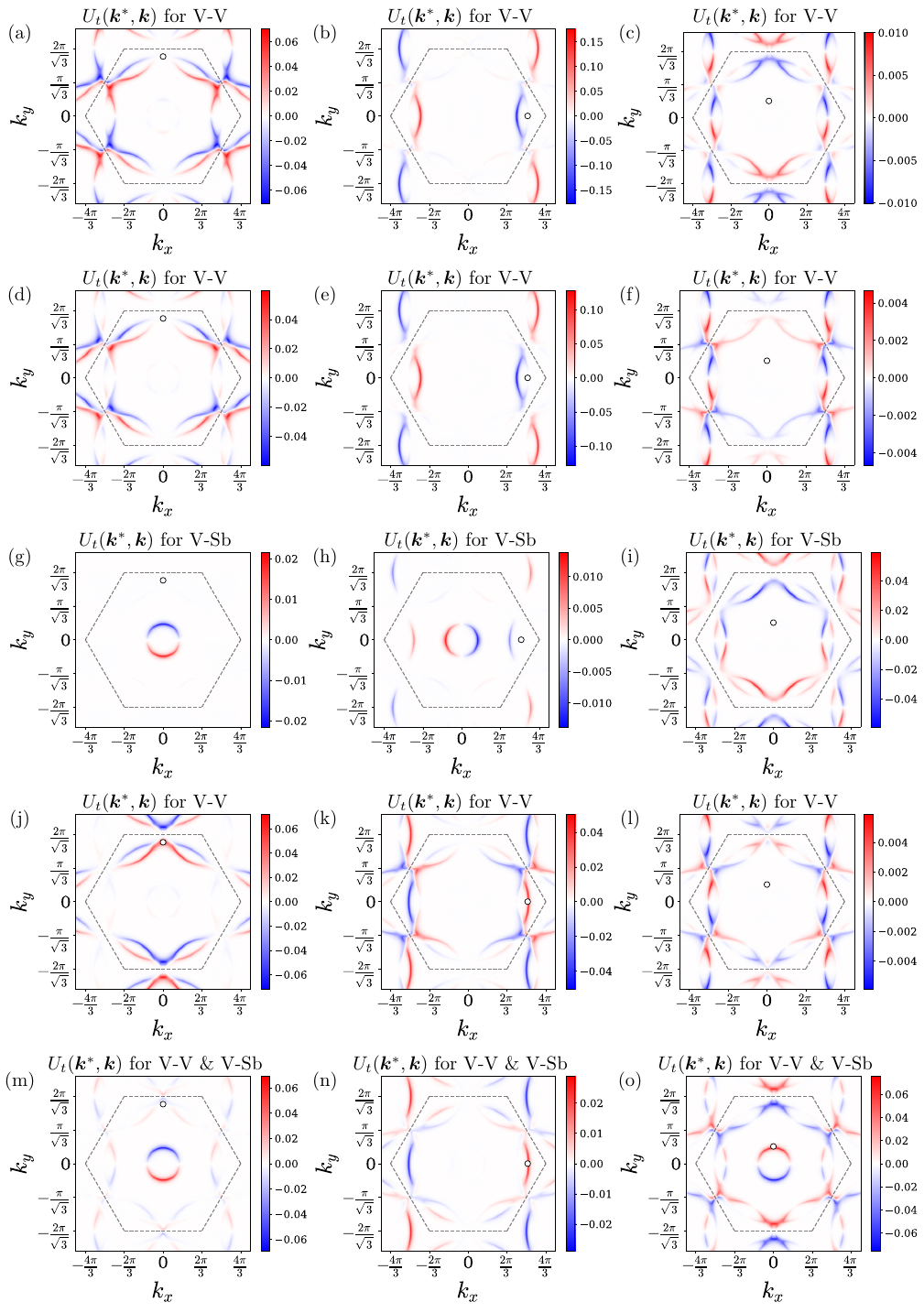}
\caption{Triplet interactions. (a)-(c) are for main text Fig.~3(a). (d)-(f) are for main text Fig.~3(b). (g)-(i) are for main text Fig.~3(c). (j)-(l) are for main text Fig.~3(d). (m)-(o) are for main text Fig.~3(e).} \label{fig:all_triplet_interactions}
\end{figure}

In this section, we show the singlet and triplet interactions generated by all current interactions. Two of these are repeated from the main text but we include them here for completeness. In Fig.~\ref{fig:all_singlet_interactions} we have all singlet interactions, and in Fig.~\ref{fig:all_triplet_interactions} we have all triplet interactions. Fig.~\ref{fig:all_singlet_interactions}(a)--(c) are the same as main text Fig.~4(a)--(c), whereas Fig.~\ref{fig:all_singlet_interactions}(g)--(i) are the same as main text Fig.~4(d)--(f). Looking at this Fig.~\ref{fig:all_singlet_interactions} for the 5 types of LC patterns, we make the following observations:

\begin{itemize}
\item If we compare the even vs.\ odd parity V-V loop current patterns, e.g. Fig.~\ref{fig:all_singlet_interactions}(a) vs.\ (j), we first see that there is still nearly no interaction to the $\Gamma$-pocket, and hence no $s^\pm$ pairing.
\item Again comparing even vs.\ odd parity V-V loop current patterns  Fig.~\ref{fig:all_singlet_interactions}(a) vs.\ (j), we see in (a) there is a strong interaction between $M$-points. However, this specific interaction is suppressed in (j) because of the odd parity of the current pattern. In spite of this, there is strong interaction to wave vectors close to the $M$-point, although not right at the $M$-point. This makes the leading pairing symmetry, $d+id$, less obvious than in the even parity case. Nonetheless, our calculations show that $d+id$ prevails in both the even and odd parity V-V LC patterns.
\item We can also compare the two different patterns which are V-Sb, one of which is parity even, and the other parity odd. In particular, if we examine Fig.~\ref{fig:all_singlet_interactions}(g) vs.\ (m), we see that there is a large interaction between the outer Fermi surface and the inner Fermi surface. This fact is true regardless of the parity, and tells us that the favoured pairing symmetry is likely to be $s^\pm$, which is indeed what is found through the linearirzed gap equation calculation.
\end{itemize}

\textit{We therefore conclude that the pairing symmetry depends fundamentally on the microscopic loop current pathway, as this strongly determines the Fermi surface regions that feel the largest interaction.}

\newpage

\bibliographystyle{apsrev4-1}

%

\end{document}